\documentclass[]{agujournal2019}
\usepackage{url} %this package should fix any errors with URLs in refs.
\usepackage{lineno}
\usepackage[inline]{trackchanges} %for better track changes. finalnew option will compile document with changes incorporated.
\usepackage{soul}

\usepackage{calc}
\usepackage{mathtools}
\usepackage{subcaption}
\usepackage{amsmath}
\usepackage{placeins}
\usepackage{amssymb}
\usepackage{algorithm}
\usepackage{algorithmic}
\DeclareMathAlphabet\mathbfcal{OMS}{cmsy}{b}{n}
\newcommand{\gt}[1]{\mathbf{\bar{#1}}} % global tensor 

\let\oldalign\align
\let\oldendalign\endalign

\renewenvironment{align}
  {\linenomath\oldalign}
  {\oldendalign\endlinenomath}

\draftfalse

\journalname{Water Resources Research}

\begin{document}

\title{Bayesian Poroelastic Aquifer Characterization from 
  \\InSAR Surface Deformation Data\\
  Part II: Quantifying the Uncertainty}

\authors{Amal Alghamdi \affil{1}, Marc A. Hesse \affil{1,2}, Jingyi Chen
\affil{2,3},  Umberto Villa\affil{4}, Omar Ghattas \affil{1,2,5}}

\affiliation{1}{University of Texas at Austin,
Oden Institute for Computational Engineering and Sciences, 
Austin, TX, United States}
 \affiliation{2}{University of Texas
at Austin, Geological Sciences, Austin, TX, United States}
\affiliation{3}{University of Texas at Austin, Aerospace
Engineering \& Engineering Mechanics, Austin, TX, United States}
\affiliation{4}{Washington University in St. Louis, Electrical and Systems Engineering, St.\ Louis, MO, United States}
\affiliation{5}{University of Texas at Austin,
Mechanical Engineering, Austin, TX, United States}

%(repeat as many times as is necessary)

%% Corresponding Author:
% Corresponding author mailing address and e-mail address:

% (include name and email addresses of the corresponding author.  More
% than one corresponding author is allowed in this LaTeX file and for
% publication; but only one corresponding author is allowed in our
% editorial system.)

% Example: \correspondingauthor{First and Last Name}{email@address.edu}

\correspondingauthor{Amal Alghamdi}{amal.m.alghamdi@gmail.com}

%% Keypoints, final entry on title page.

%  List up to three key points (at least one is required)
%  Key Points summarize the main points and conclusions of the article
%  Each must be 100 characters or less with no special characters or punctuation and must be complete sentences

\begin{keypoints}

\item Using InSAR data reduces the uncertainty in selected quantities of interest compared to using prior knowledge only
\item The preconditioned Crank--Nicolson (pCN) MCMC method is extended to exploit posterior curvature and allow better chain mixing 
\item We demonstrate the intrinsic low dimensionality of the poroelastic inverse problem that is critical for the success of the MCMC method 

\end{keypoints}

%% ------------------------------------------------------------------------ %%
%
%  ABSTRACT and PLAIN LANGUAGE SUMMARY
%
% A good Abstract will begin with a short description of the problem
% being addressed, briefly describe the new data or analyses, then
% briefly states the main conclusion(s) and how they are supported and
% uncertainties.

% The Plain Language Summary should be written for a broad audience,
% including journalists and the science-interested public, that will not have 
% a background in your field.
%
% A Plain Language Summary is required in GRL, JGR: Planets, JGR: Biogeosciences,
% JGR: Oceans, G-Cubed, Reviews of Geophysics, and JAMES.
% see http://sharingscience.agu.org/creating-plain-language-summary/)
%
%% ------------------------------------------------------------------------ %%
%% \begin{abstract} starts the second page

\begin{abstract}
%Abstract should be one paragraph and should be less than 250 word
Uncertainty quantification of groundwater (GW) aquifer parameters is critical for efficient management and sustainable extraction of GW resources. These uncertainties are introduced by the data, model, and prior information on the parameters. Here we develop a Bayesian inversion framework that uses Interferometric Synthetic Aperture Radar (InSAR) surface deformation data to infer the laterally heterogeneous permeability of a transient linear poroelastic model of a confined GW aquifer. The Bayesian solution of this inverse problem takes the form of a posterior probability density of the permeability. Exploring this posterior using classical Markov chain Monte Carlo (MCMC) methods is computationally prohibitive due to the large dimension of the discretized permeability field and the expense of solving the  poroelastic forward problem. However, in many partial differential equation (PDE)-based Bayesian inversion problems, the data are only informative in a few directions in parameter space. For the poroelasticity problem, we prove this property theoretically for a one-dimensional problem and demonstrate it numerically for a three-dimensional  aquifer model. We design a generalized preconditioned Crank--Nicolson (gpCN) MCMC method that exploits this intrinsic low dimensionality by using a low-rank based Laplace approximation of the posterior as a proposal, which we build scalably. The feasibility of our approach is demonstrated through  a real GW aquifer test in Nevada. The inherently two dimensional nature of InSAR surface deformation data informs a sufficient number of modes of the permeability field to allow detection of major structures within the aquifer, significantly reducing the uncertainty in the pressure and the displacement quantities of interest.

\end{abstract}

%%%%%%%%%%%%%%%%%%%%%%%%%%%%%%%%%%%%%%%%%%%%%%%%%%%%%%%%%%%%%%%%%%%%%%%%%%%%%%%
%%%%%%%%%%%%%%%%%%%%%%%%% Section: Introduction %%%%%%%%%%%%%%%%%%%%%%%%%%%%%%%
%%%%%%%%%%%%%%%%%%%%%%%%%%%%%%%%%%%%%%%%%%%%%%%%%%%%%%%%%%%%%%%%%%%%%%%%%%%%%%%
\section{Introduction}
\label{sec: Intro}

% #1 context and overall goal
Efficient groundwater (GW) resources management is critical for mitigating the global-scale problem of GW depletion \cite{wadaVanBeekVanKempenEtAl2010,wadaVanBeekBierkens2012,wadaWisserBierkens2014,famiglietti2014}. This is the result of extracting GW at rates that exceed the natural recharge, which then leads to significant drops in GW pressure and (possibly irreversible) compaction \cite{wadaVanBeekVanKempenEtAl2010,konikowKendy2005,xueZhangYeEtAl2005,holzerGalloway2005}.  Making informed management and control decisions requires accurate GW aquifer models, typically in the form of a time-dependent partial differential equation (PDE). These models are used to predict the aquifer response (e.g.\ GW pressure drop and aquifer compaction) to GW extraction activities.  The major source of uncertainty in these models is the uncertainty in the aquifer properties  \cite{Eaton2006,Bohling2010,Oliver2011}. The goal of our work is to devise a framework to systematically quantify the uncertainty in GW aquifer model properties. Here we only consider uncertainty in one parameter field, the aquifer permeability, and one data source, surface deformation data, but the framework we present can be generalized to multiple parameter fields and data sources. To estimate the permeability, we take advantage, in particular, of the evergrowing InSAR data that provide  frequent, large-scale, and up to sub-millimeter accurate surface deformation measurements globally \cite{FerrettiSavioBarzaghiEtAL2007,TomasRomeroMulas2014}.

% #2 Problem, challenges and need
Bayes' theorem---which has been adopted in GW applications since as early as 1986 \cite{Carrera1986a,Carrera1986b,Carrera1986c,McLaughlin1996,LindeGinsbourgerIrvingEtAl2017}---provides a natural means of incorporating observational data, e.g., pressure or surface deformation data at the aquifer site, to quantify the uncertainty in the aquifer model parameters (such as physical properties, source terms, and/or boundary or initial conditions). The Bayesian framework updates our assumed ``prior" knowledge about the aquifer parameters through what is known as the ``likelihood" distribution, which determines how likely it is for the observational data to result  from the modeled system with a particular parameter realization.  The updated probability distribution is called the ``posterior" distribution and is regarded as the solution of the inverse problem. Bayesian inversion, therefore, provides a characterization of the uncertainty in the parameters, as opposed to finding just a point estimate as is with deterministic inversion.

The Markov chain Monte Carlo (MCMC) method is often the method of choice for characterizing the posterior distribution for PDE-based Bayesian inverse problems.  MCMC generates samples of the posterior probability density function (PDF)---each of which requires solution of at least the forward PDEs---from which sample statistics can be computed.  Conventional MCMC methods view the parameter-to-observable map as a black box and thus are not capable of exploiting the structure of this map to accelerate convergence of MCMC chains.  As such, most work on Bayesian inversion governed by PDEs has been restricted to simple PDEs or low parameter dimensions (or both). 

Over the past decade, several powerful MCMC methods have emerged that exploit the geometry and smoothness---and in some cases the underlying low-dimensionality---of the parameter-to-observable map to enable efficient MCMC sampling from high-dimensional PDE-based posteriors \cite{CuiLawMarzouk16,FlathWilcoxAkcelikEtAl11,MartinWilcoxBursteddeEtAl12, Bui-ThanhGhattasMartinEtAl13,PetraMartinStadlerEtAl14,BeskosGirolamiLanEtAl17}. These methods have been applied to several large-scale Bayesian inverse problems governed by complex forward problems, including seismic wave propagation  \cite{MartinWilcoxBursteddeEtAl12} and ice sheet flow \cite{IsaacPetraStadlerEtAl15}. What these methods all have in common is the use of the Hessian of the negative log posterior to exploit the geometry of the posterior. Moreover, the low rank structure of the Hessian can be used to identify the (often) low dimensional manifold (of dimension $r$ much smaller than the nominal dimension of the discretized parameters) on which the parameters are informed by the data. To that end, randomized  methods (e.g., randomized singular value decomposition and randomized generalized eigensolvers) are devised to identify this low dimensional manifold at a cost of $O(r)$ PDE solves only. Characterizing high-dimensional posterior distributions governed by three-dimensional transient poroelastic GW aquifer models has been explored by \citeA{HesseStadler14} via a Gaussian approximation of the posterior distribution for model problems. Full MCMC sampling of posteriors governed by high-dimensional GW models of such complexity has yet to be explored.

% #3 Contribution 
In \citeA{AlghamdiHesseChenEtAl2020A}, Part I of this two-part series of articles, we built a Bayesian framework to characterize the lateral heterogeneity in a three-dimensional time-dependent poroelastic aquifer model and applied it to a test case in Nevada using InSAR surface deformation data. There, we addressed the choice of the prior distribution and the data noise model. We also took a first step in exploring the posterior by inverting for the maximum a posteriori (MAP) point, which is the most likely permeability realization with respect to the posterior distribution. We used independent GPS measurements \cite{BurbeyWarnerBlewittEtAl06} for validation. A brief summary of the pertinent results from Part I is provided in section~\ref{sec: Nevada Test Case}. 

We summarize the contribution of Part II  as follows: (1) We demonstrate the decaying eigenvalues of the data misfit part of the Hessian analytically, through Fourier analysis, for a simplified 1D poroelasticity problem. We additionally provide numerical evidence of the data misfit Hessian compactness for the three-dimensional fully-coupled quasi-static linear poroelastic aquifer model considered here. This property determines the applicability of the aforementioned class of scalable MCMC methods. (2) We propose a generalization of the preconditioned Crank--Nicolson (pCN) MCMC method \cite{CotterRobertsStuartEtAl12}, referred to as gpCN thereafter,  that identifies and exploits this low dimensional manifold using randomized eigensolvers to construct a low-rank based Laplace approximation of the posterior. Besides the initial cost of creating the low-rank based Laplace approximation, which is negligible in comparison to the cost of generating MCMC chains, gpCN has no additional cost per MCMC sample in comparison to pCN. Furthermore, we compare convergence properties of pCN and gpCN. (3) We apply gpCN to characterize the high-dimensional subsurface permeability for the Nevada test case aquifer, and quantify the associated uncertainty (both in the parameters and the selected state-variable-based quantities of interest (QoIs)). We discuss how  information from the data reduces the prior uncertainty in the parameters and the state variables in terms of  the location in the physical domain and the dominant parameters modes inferred from the data. The three integrated contributions demonstrate a successful design and application of a scalable MCMC method to a real-world GW Bayesian inverse problem governed by a  coupled forward problem with high dimensional parameters and using InSAR data.

% #4 outline
The reminder of the paper is structured as follows. In section \ref{sec: methods}, we review the quasi-static linear poroelasticity model and the Bayesian framework that we established in Part I. We discuss low-rank based Laplace approximation of the posterior and present gpCN, our generalization of pCN MCMC method. We describe the test case to which we apply our framework in section \ref{sec: Nevada Test Case}. We demonstrate gpCN performance and present results on characterizing the uncertainty in the permeability and selected state-variable-based QoI for the Nevada test case  in section~\ref{sec: Results and Discussion}. We summarize our conclusions in section~\ref{sec: Conclusion}.

\section{Methods}
\label{sec: methods}

In Part I of this two-part series of articles, we formulate a Bayesian inversion framework governed by quasi-static linear poroelasticity. The solution of the Bayesian inverse problem takes the form of a posterior distribution that characterizes the heterogeneity of groundwater aquifer permeability. Here, we review the forward model, the quasi-static linear poroelasticity model, and the formulation of the inverse problem, the Bayesian framework, in sections \ref{sec: Poroelasticity Background} and \ref{sec: Bayesian Framework Background}, respectively. We also demonstrate the compactness of the data misfit Hessian for a simplified one dimensional poroelasticity problem.  In sections \ref{sec: Laplace approximation} and \ref{sec: gpCN}, we build a low-rank based Laplace approximation and present the gpCN MCMC method that exploits this intrinsic low dimensionality in the poroelasticity inverse problem.

\subsection{Quasi-Static Linear Poroelasticity}
\label{sec: Poroelasticity Background}

In the Bayesian inversion framework, we model the aquifer as a poroelastic medium governed by the linear poroelasticity theory developed by \citeA{Biot1941}. Biot theory describes the coupling between the fluid pressure field in a saturated porous medium and the accompanying elastic deformation of the solid skeleton. The classical formulation of Biot system in a space-time domain $\Omega\times(0,T]$ can be written as % assuming well-defined boundary and initial conditions:
    \begin{align}
    \left(S_\epsilon p + \alpha \nabla\cdot\mathbf{u} \right)_t - \nabla \cdot\left(\frac{\kappa}{\mu}\nabla p\right) =& f_p \nonumber \\
    -\nabla\cdot\left(\pmb{\sigma}(\mathbf{u}) - \alpha p \mathbf{I} \right) =& \mathbf{f_u},  \label{equ: Biot model b}
    \end{align}
where $(\cdot)_t$ denotes time derivative,  $p=p(\mathbf{x},t)$ is the deviation from hydrostatic pressure,  $\mathbf{u}=\mathbf{u}(\mathbf{x},t)$ is the displacement of the solid skeleton, $f_p =f_p(\mathbf{x},t)$ is a fluid source per unit volume, $\mathbf{f_u} =\mathbf{f_u}(\mathbf{x},t)$ is the body force per unit volume,   $S_\epsilon$ is the specific storage, $\kappa(\mathbf{x})$ is the medium permeability field, $\mu$ is the dynamic viscosity of the pore fluid, and $\alpha$ is the Biot--Willis coupling parameter. The tensor $\pmb{\sigma}(\mathbf{u})$ is the stress tensor for isotropic linear elasticity, and it is parameterized by the drained shear modulus $G$ and the Poisson's ratio $\nu$. We write the unknown medium permeability $\kappa(\mathbf{x})$ in terms of the log-permeability field, which we denote $m(\mathbf{x})$, so that $\kappa(\mathbf{x})= e^{m(\mathbf{x})}$. This parameterization ensures that the inferred permeability field is positive when solving the inverse problem for $m(\mathbf{x})$. The system \eqref{equ: Biot model b} together with well-defined boundary and initial conditions, given in  \citeA{AlghamdiHesseChenEtAl2020A}, form the forward problem.

We use a three-field formulation of the poroelasticity system~\eqref{equ: Biot model b} in which  Darcy flux is introduced as a state variable $\mathbf{q}(\mathbf{x},t) =-\frac{\kappa}{\mu}\nabla p$, adding a third equation to the system~\eqref{equ: Biot model b}. Following \citeA{FerronatoCastellettoGambolati10}, we employ a mixed finite element method (MFEM) to discretize the three-field formulation in space and use backward Euler method for discretization in time. In this MFEM,  the pressure $p$ is approximated by piecewise constant functions, the displacement $\mathbf{u}$ is approximated in the first-order  Lagrange polynomial space, and the flux $\mathbf{q}$ is approximated in the lowest-order Raviart--Thomas space. This MFEM discretization has the advantage of conserving mass discretely and reducing the non-physical pressure oscillations that can form in some other discretizations of the poroelasticity system \cite{PhillipsWheeler2009,FerronatoCastellettoGambolati10,HagaOsnesLangtangen12}.  To simplify notation, we define the discretized ``global'' space-time system, which combines solving for all time steps simultaneously, as 
\begin{align}
	\gt{S}(\mathbf{m})\gt{X} = \gt{F}, \label{equ: Global Time Space System}
\end{align}
where the block lower triangular matrix $\gt{S}$ combines the discretized differential operators of the three-field formulation of the poroelasticity system~\eqref{equ: Biot model b}. $\gt{S}$ depends on the discretized log permeability  $\mathbf{m}\in \mathbb{R}^{n_n}$, which we approximate in the first order Lagrange polynomial space, where $n_n$ is the number of degrees of freedom (DOFs) of the discretized parameter field. The vector $\gt{F}$ combines the source terms and the natural boundary conditions at all time steps, and the vector $\gt{X}$ consists of the pressure, displacement, and fluid flux DOFs at all time steps as well. The discretization details can be found in Part I  \cite{AlghamdiHesseChenEtAl2020A}.

\subsection{Poroelasticity Bayesian Inversion Framework Constrained by InSAR Surface Deformation Data}
\label{sec: Bayesian Framework Background}

Our goal is to characterize the aquifer heterogeneous permeability $\kappa=e^m$ in the poroelasticity model~\eqref{equ: Biot model b} from InSAR  Line-of-Sight (LOS) surface deformation data $\mathbf{d}^{\text{obs}}  \in \mathbb{R}^{n_\text{obs}}$, where $n_\text{obs}$ is the number of observational data (total number of pixels in deformation maps in this case). The LOS surface deformation $u_\text{LOS}$ is given by
\begin{align}
u_\text{LOS}= \alpha_1 u_1 +  \alpha_2 u_2 +  \alpha_3 u_3,
\label{equ: los}
\end{align}
where $u_1$, $u_2$ and $u_3$ are the east, north and vertical surface displacements, respectively, and $\alpha_1$, $\alpha_2$ and $\alpha_3$ are the eastward, northward and vertical components of the LOS unit look vector $\alpha_\text{LOS}= [\alpha_1, \alpha_2, \alpha_3]$. In this study, we only use one LOS net deformation map available at $t=t_\text{InSAR}$. In general, multiple deformation maps or deformation time series can be available.

Assuming modeling errors are negligible compared to noise in the data, for a given log permeability field $\mathbf{m}(\mathbf{x})$, the relation between the observational data $\mathbf{d}^{\text{obs}}$ and the model output is given by 
\begin{align}
\mathbf{d}^{\text{obs}} = \mathbfcal{F}(\mathbf{m}) + \pmb\eta \label{equ: Additive Noise Form}.
\end{align}
The map $\mathbfcal{F}(\cdot):   \mathbb{R}^{n_n} \rightarrow  \mathbb{R}^{n_{\text{obs}}}  $ is the  \textit{parameter-to-observable map} which, for a given $\mathbf{m}$, gives the observables, i.e\ the modeled LOS surface deformation values at the observational data locations (pixels centers) and times. Evaluating $\mathbfcal{F}(\mathbf{m})$ requires solving model~\eqref{equ: Biot model b} using the given log permeability $\mathbf{m}$. 
In particular, $\mathbfcal{F}(\mathbf{m})\coloneqq \gt{B}\gt{X} = \gt{B}\gt{S}(\mathbf{m})^{-1}\gt{F}$, where we define the pointwise linear observation operator $\gt{B}$ which extracts the  pointwise observables from the modeled displacement DOFs included in the global vector $\gt{X}$. The solution $\gt{X}$ is obtained by solving the ``global'' system~\eqref{equ: Global Time Space System}.
The random vector  $\pmb\eta \sim \mathcal{N}(\mathbf{0}, \mathbf{\Gamma}_\text{noise})$ is measurement noise which we assume is additive and Gaussian with mean $\mathbf{0}$ and covariance matrix $\mathbf{\Gamma}_\text{noise}  \in  \mathbb{R}^{n_\text{obs}\times n_\text{obs} }$. 

We use relation~\eqref{equ: Additive Noise Form} to define the likelihood probability distribution that determines how likely it is for the observational data $\mathbf{d}^{\text{obs}}$ to be measured from an aquifer with permeability $\mathbf{m}$---in the absence of model error. In other words, the likelihood distribution is the probability distribution of the discrepancy between the observed data and the model outcome which, based on the relation~\eqref{equ: Additive Noise Form}, is given by
\begin{align}
  \pi_\text{likelihood}(\mathbf{d}^\text{obs} | \mathbf{m})  \propto  e^{-\ell(
 \mathbf{d}^\text{obs};\mathbf{m})},
\end{align}
where the symbol $\propto$ denotes equality up to a multiplicative constant and $\ell$ is the negative log of the likelihood, or the data misfit, which is given by
\begin{align}
\ell( \mathbf{d}^\text{obs}; \mathbf{m}) = \frac{1}{2}(  \mathbfcal{F}(\mathbf{m}) -
\mathbf{d}^{\text{obs}})^T\mathbf{\Gamma}_\text{noise}^{-1}(   \mathbfcal{F}(\mathbf{m}) -
\mathbf{d}^{\text{obs}} ).
\label{equ: Negative Log Liklihood}
\end{align}

Bayes theory gives the posterior probability distribution $\pi_\text{post}(\mathbf{m}| \mathbf{d}^\text{obs})$, which quantifies the uncertainty in the parameter $\mathbf{m}$ given the observational data $\mathbf{d}^{\text{obs}}$. The posterior distribution  incorporates information about the model and the data, via the likelihood distribution. Statistical prior assumptions about the parameters are encoded into the prior distribution $\pi_\text{prior}(\mathbf{m})$. Bayesian formulation of inverse problems was introduced in \cite{TarantolaValette1982} and is given by
\begin{align}
\pi_\text{post}(\mathbf{m}| \mathbf{d}^\text{obs}) &\propto \pi_\text{likelihood}(\mathbf{d}^\text{obs} | \mathbf{m}) \pi_\text{prior}(\mathbf{m}) \nonumber \\
%Thus we can rewrite the the posterior distribution~\refp{equ: Bayes Formula Finite} as:
& \propto   \text{exp}\left(- \frac{1}{2}(
\gt{B}\gt{X} - \mathbf{d}^{\text{obs}}
)^T\mathbf{\Gamma}_\text{noise}^{-1} ( \gt{B}\gt{X} -
\mathbf{d}^{\text{obs}}  )  -  \frac{1}{2}(\mathbf{m} -
\bar{\mathbf{m}})^T\mathbf{\Gamma}_\text{prior}^{-1} (\mathbf{m} -
\bar{\mathbf{m}})\right). 
\label{equ: Bayes Formula Finite}
\end{align} 
We assume the prior distribution is a discretized Mat\'ern Gaussian field with mean $\bar{\mathbf{m}} \in  \mathbb{R}^{n_n}$ and covariance matrix $\mathbf{\Gamma}_\text{prior}  \in  \mathbb{R}^{n_n\times n_n }$. The latter dictates our assumptions about the pointwise variance and the spatial correlation of the log permeability field features. For this class of priors, the precision operator $\mathbf{\Gamma}_\text{prior}^{-1}$ admits a representation $\mathbf{\Gamma}_\text{prior}^{-1}= \mathbf{A}\mathbf{M}_{m}^{-1}\mathbf{A}$ in terms of an elliptic differential operator $\mathbf{A}$ and a mass matrix $\mathbf{M}_{m}$ in the discretized parameter space \cite{LindgrenRueLindstroem11,AlghamdiHesseChenEtAl2020A}. %For details about defining the likelihood and the prior, please refer to Part~I \cite{AlghamdiHesseChenEtAl2020A}.

The posterior distribution is a distribution in a very large dimensional space $\mathbb{R}^{n_n}$, and evaluating the posterior at a single realization $\mathbf{m}$ requires solving the model~\eqref{equ: Biot model b} forward in time. Computing moments and other expected values of interest from the posterior distribution is computationally prohibitive using traditional MCMC sampling methods. Fortunately, in many PDE-based Bayesian inversion problems, the data are informative in only relatively few directions in parameter space $r\ll n_n$. This property can be exploited to obtain low-rank based approximations of the posterior, which additionally can be used to build effective MCMC proposals, as we present in the remaining of this section.

%%%%%%%%%%%%%%%%%%%%%%%%%%%%%%%%%%%%%%%%%%%%%%%%%%%%%%%%%%%%%%%%%%%%%%%%%%%%%%%
%%%%%%%%%%%%%%%%%%%%%%%%%%%%Laplace approximation%%%%%%%%%%%%%%%%%%%%%%%%%%%%%%
%%%%%%%%%%%%%%%%%%%%%%%%%%%%%%%%%%%%%%%%%%%%%%%%%%%%%%%%%%%%%%%%%%%%%%%%%%%%%%%
\subsection{Low-Rank Based Laplace Approximation of the Posterior}
\label{sec: Laplace approximation}

\subsubsection{The Laplace Approximation}

In Part I we focused on solution of the optimization problem of maximizing the posterior distribution to find the MAP point $\mathbf{m}_\text{MAP}$. Besides being the most likely realization of the permeability field, the MAP point is also needed to form the Laplace approximation of the posterior, which is a Gaussian approximation obtained by making a quadratic approximation of the negative log posterior at the MAP point \cite{MacKay2003}. This approximation is valid if the parameter-to-observable map is approximately linear over the support of the prior distribution. Additionally, it can be used in building effective proposals for dimension invariant MCMC algorithms as we discuss in section~\ref{sec: gpCN}. 

The negative log of the posterior~\eqref{equ: Bayes Formula Finite} can be written as
\begin{align}
-\log&\left( \pi_\text{post}\left( \mathbf{m}\right) \right) \nonumber\\
  =     &\frac{1}{2}\left(\mathbfcal{F}(\mathbf{m}) - \mathbf{d}^\text{obs}\right)^T \mathbf{\Gamma}_\text{noise}^{-1} \left(\mathbfcal{F}(\mathbf{m})- \mathbf{d}^\text{obs} \right) 
       +\frac{1}{2}\left(\mathbf{m} - \bar{\mathbf{m}}    \right)^T \mathbf{\Gamma}_\text{prior}^{-1}\left( \mathbf{m} - \bar{\mathbf{m}}    \right) - c_1, \label{equ: Taylor Expansion a}
\end{align}
where $c_1$ is the log of the posterior normalization constant.  Using a Taylor expansion  at the point $\mathbf{m}_\text{MAP}$ in  parameter space, a quadratic approximation of \eqref{equ: Taylor Expansion a} is given by
\begin{align}
-\log&\left( \pi_\text{post}\left( \mathbf{m}_\text{MAP} + \hat{\mathbf{m}}  \right)\right) \approx c_2+\frac{1}{2}{\hat{\mathbf{m}}}^T \mathbf{H}(\mathbf{m}_\text{MAP})  \hat{\mathbf{m}} ,
\label{equ: Taylor Expansion b}
\end{align}
where $\hat{\mathbf{m}} = \mathbf{m}-\mathbf{m}_\text{MAP}$. To obtain~\eqref{equ: Taylor Expansion b}, we use the fact that the gradient  of the negative log posterior with respect to the parameter $\mathbf{m}$ is zero at the MAP point. The constant $c_2$ is defined as
\begin{align}
c_2 = &\frac{1}{2}\left(\mathbfcal{F}(\mathbf{m}_\text{MAP}) - \mathbf{d}^\text{obs} \right)^T \mathbf{\Gamma}_\text{noise}^{-1} \left(\mathbfcal{F}(\mathbf{m}_\text{MAP})-\mathbf{d}^\text{obs} \right) \nonumber\\
        +&\frac{1}{2}\left(\mathbf{m}_\text{MAP}  - \bar{\mathbf{m}}    \right)^T \mathbf{\Gamma}_\text{prior}^{-1}\left( \mathbf{m}_\text{MAP}  - \bar{\mathbf{m}}    \right) - c_1. 
\label{equ: Constant for Taylor Expansion}
\end{align}
 The matrix $\mathbf{H}(\mathbf{m}_\text{MAP}) \in  \mathbb{R}^{n_n\times n_n }$ is the Hessian operator of the negative log of the posterior~\eqref{equ: Bayes Formula Finite} evaluated at the MAP point and it is given by
\begin{align}
\mathbf{H}(\mathbf{m}) = \mathbf{H}_\text{misfit}(\mathbf{m}) + \mathbf{\Gamma}_\text{prior}^{-1},
\end{align}
where the data misfit Hessian $\mathbf{H}_\text{misfit}(\mathbf{m})$ is given by 
\begin{align}
\mathbf{H}_\text{misfit}(\mathbf{m}) = \nabla\mathbfcal{F}(\mathbf{m})^T\mathbf{\Gamma}_\text{noise}^{-1}\nabla\mathbfcal{F}(\mathbf{m}) + \sum_i^{n_\text{obs}} {\left(\mathbf{\Gamma}_\text{noise}^{-1} (\mathbfcal{F}(\mathbf{m}) - \mathbf{d}^\text{obs}) \right)_i \nabla^2 \mathbfcal{F}_i(\mathbf{m})}.
\end{align}
The resulting Laplace approximation is the Gaussian distribution:
\begin{align}
 \pi_\text{Laplace}(\mathbf{m}) \sim & \mathcal{N}\left(\mathbf{m}_\text{MAP}, \mathbf{H}(\mathbf{m}_\text{MAP})^{-1}\right).
\label{Gaussian_around_the_map}
\end{align}

\subsubsection{Spectral Properties of the Hessian}

In many PDE-based inverse problems, it is often the case that the data misfit Hessian operator is compact and only a few parameter dimensions, $O(r)$, are informed by the data \cite{FlathWilcoxAkcelikEtAl11,Bui-ThanhBursteddeGhattasEtAl12,IsaacPetraStadlerEtAl15,VillaPetraGhattas2020}. In \citeA{Alghamdi2020}, we provide an analytical formula for the eigenvalues and the eigenfunctions of the Hessian of an idealized one-dimensional steady-state poroelasticity problem. We assume this problem is defined on an interval $[0,L]$, and that we have continuous deformation observations everywhere in this interval. Specifically, the eigenvalues $\lambda_i$ and eigenfunctions $\phi_i$ of the data misfit Hessian operator in this case are given by
\begin{align}
    \lambda_i = \frac{ f_L^2 \mu^2}{e^2\kappa_0^4} \left(\frac{2L}{i \pi}\right)^4, \quad \phi_i(x) = \cos\left(\frac{i \pi x}{2 L}\right), \quad i = 1,3,5,...,
    \label{equ: Eigenfunctions}
\end{align}
where $f_L$ is the Darcy flux at the boundary $x=L$, $e$ is Young's modulus, and $\kappa_0$ is the constant permeability value at which the data misfit Hessian operator is evaluated. We note from the expressions in~\eqref{equ: Eigenfunctions} that the eigenvalues have a rapid decay of $O(\frac{1}{i^4})$, and the eigenfunctions  corresponding to dominant eigenvalues are smoother. This means that the more oscillatory a permeability mode is, the more difficult it is to infer from the deformation data. Note that the expected information gain (from prior to posterior) from the data in each mode is given by $\log (1 + \lambda_i)$ under a Gaussian assumption \cite{AlexanderianGloorGhattas16}. The rapid decay of eigenvalues in ~\eqref{equ: Eigenfunctions} implies that the data inform the permeability field in a low dimensional manifold. Although this analysis is carried out for an idealized one-dimensional case, numerical evidence suggests  rapid eigenvalue decay---independent of the discretization dimension---of the  data misfit Hessian of the 3D poroelasticity problem we study; see section~\ref{sec: Results and Discussion}. Moreover, analogous to the 1D analysis result, larger eigenvalues correspond to smoother eigenvectors.

\subsubsection{Sampling from the Laplace Approximation}

The gpCN method described in the next section requires the ability to generate samples $\pmb\eta_\text{Laplace}$ from the distribution $ \pi_\text{Laplace}(\mathbf{m})$.
To sample from this Gaussian distribution, $\mathbf{H}^{-\frac{1}{2}}$ is applied to a white noise vector.  However, constructing the Hessian matrix explicitly and   computing its Cholesky factorization is computationally prohibitive in high parameter dimensions. The Hessian construction costs $n_n$ pairs of incremental forward \eqref{equ: Incremental Forward} and incremental adjoint \eqref{equ: Incremental Adjoint} poroelasticity solves; see \ref{app: Hessian} for details.

Instead of building the Hessian explicitly, we take advantage of the fact that applying the Hessian to a vector can be carried out in a matrix-free manner and costs only a pair of incremental forward and incremental adjoint solves, and the data misfit Hessian is inherently low-rank. This allows us to build a low-rank based representation of the Hessian at $\mathbf{m}_\text{MAP}$ by solving a generalized eigenvalue problem using a randomized generalized eigensolver---at a cost of just $O(r)$ incremental forward and adjoint poroelasticity solves. We use this approximation to create a low-rank based Laplace approximation, $\pi_\text{Laplace,r}(\mathbf{m})$, of the posterior.

The following approach of approximating the Hessian at the MAP point has been applied to the inverse problem for the basal friction coefficient of the Antarctic ice sheet
\cite{IsaacPetraStadlerEtAl15}, and is integrated in the \texttt{hIPPYlib} library \cite{VillaPetraGhattas18}. This approach is based on solving the generalized eigenvalue problem:
 \begin{align}
\mathbf{H}_\text{misfit}(\mathbf{m}_\text{MAP})\mathbf{w}_i= \lambda_i \mathbf{\Gamma}_\text{prior}^{-1}\mathbf{w}_i  \quad
\text{for}\quad i=1,..,r.
\label{equ: Generalized Eigenvalue Problem}
\end{align}
 The eigenvalues $\lambda_i$ of this generalized eigenvalue problem  indicate whether the corresponding eigendirection $\mathbf{w}_i$  in the high dimensional parameter space $\mathbb{R}^{n_n}$ is mostly informed by the data (when $\lambda_i>1$) or by the prior (when $\lambda_i<1$). Also, the magnitude of $\lambda_i$ reflects the level of certainty of inferring the corresponding eigendirection from the data. The diagonal matrix of the first $r$ dominant eigenvalues, $\lambda_i$, and the matrix of the corresponding eigenvectors, $\mathbf{w}_i$,  of this generalized eigenvalue problem are denoted as $\mathbf{\Lambda}_r = \text{diag}([\lambda_1,\lambda_2, ..., \lambda_r] )$ and  $\mathbf{W}_r = [\mathbf{w}_1, .., \mathbf{w}_r ]$, respectively. The matrix $\mathbf{W}_r$ has $\mathbf{\Gamma}_\text{prior}^{-1}$-orthonormal columns, that is  $\mathbf{W}_r^T\mathbf{\Gamma}_\text{prior}^{-1}\mathbf{W}_r= \mathbf{I}_r$, where $\mathbf{I}_r$ denotes the identity matrix in $\mathbb{R}^{r}$.
 
 The eigenvalue problem~\eqref{equ: Generalized Eigenvalue Problem} along with the Sherman--Woodbury formula are then used to obtain the following low-rank approximation of the Laplace approximation covariance matrix (the inverse of the Hessian at the MAP point); see \cite{VillaPetraGhattas2020}: 
\begin{align}
\mathbf{H}(\mathbf{m}_\text{MAP})^{-1} &=  ( \mathbf{H}_\text{misfit}(\mathbf{m}_\text{MAP})+\mathbf{\Gamma}_\text{prior}^{-1})^{-1}\nonumber\\
& \approx  \mathbf{\Gamma}_\text{prior} -  \mathbf{W}_r\mathbf{D}_r \mathbf{W}_r^T.
\label{equ: Low Rank Based Lapalce Approximation}
\end{align}
The matrix $\mathbf{D}_r$ is given by $\mathbf{D}_r = \text{diag}([\frac{\lambda_1}{1+\lambda_1},\frac{\lambda_2}{1+\lambda_2},...,\frac{\lambda_r}{1+\lambda_r}])$. The value $\frac{\lambda_i}{1+\lambda_i}$ is negligible for $\lambda_i \ll 1$, which provides a basis for truncation. We denote the low-rank based covariance matrix given by the inverse of the Hessian in~\eqref{equ: Low Rank Based Lapalce Approximation} by  $\mathbf{\Gamma}_\text{Laplace,r} \coloneqq \mathbf{\Gamma}_\text{prior} -  \mathbf{W}_r\mathbf{D}_r \mathbf{W}_r^T$. A sample $\pmb\eta_\text{Laplace,r}$ from this low-rank based Laplace approximation $\pi_\text{Laplace,r}(\mathbf{m})= \mathcal{N}(\mathbf{m}_\text{MAP}, \mathbf{\Gamma}_\text{Laplace,r} )$ can be  generated scalably using \cite{VillaPetraGhattas2020}:
\begin{align}
\pmb\eta_\text{Laplace,r} = (\mathbf{I} -  \mathbf{W}_r\mathbf{S}_r \mathbf{W}_r^T\mathbf{\Gamma}_\text{prior}^{-1})\pmb\eta + \mathbf{m}_\text{MAP}
\end{align}
where $\mathbf{S}_r = \text{diag}([1- \frac{1}{\sqrt{1+\lambda_1}}, 1- \frac{1}{\sqrt{1+\lambda_2}}, ..., 1- \frac{1}{\sqrt{1+\lambda_r}}])$ and $\pmb\eta \sim \mathcal{N}(0,\mathbf{\Gamma}_\text{prior})$. The cost of sampling from this approximation is dominated by the cost of generating the prior sample $\pmb\eta$ (mainly applying $\mathbf{A}^{-1}$ via a scalable multigrid solve as discussed in Part I), and the matrix--vector multiplications by sparse and low-rank matrices.

%%%%%%%%%%%%%%%%%%%%%%%%%%%%%%%%%%%%%%%%%%%%%%%%%%%%%%%%%%%%%%%%%%%%%%%%%%%%%%%
%%%%%%%%%%%%%%%%%%%%%%%Dimension invariant MCMC methods%%%%%%%%%%%%%%%%%%%%%%%%
%%%%%%%%%%%%%%%%%%%%%%%%%%%%%%%%%%%%%%%%%%%%%%%%%%%%%%%%%%%%%%%%%%%%%%%%%%%%%%%
\subsection{Generalized Preconditioned Crank-Nicolson (gpCN) MCMC method}
\label{sec: gpCN}

To characterize the posterior distribution,  MCMC methods generate samples of the posterior PDF (each of which requires at least a forward PDE solve) from which sample statistics can be computed. As stated earlier, for PDE-governed posteriors in a high dimensional parameter space, it is often the case that the low rank structure of the Hessian can be used to identify the low dimensional manifold on which the parameters are informed by the data, which limits the expensive PDE solves to exploration of just a low dimensional subspace of the parameter space.  

In this work, we design a generalization of the preconditioned Crank--Nicolson (gpCN) MCMC method. Preconditioned Crank--Nicolson (pCN)  is a dimension-invariant MCMC method that uses prior information to build the proposal distribution  \cite{CotterRobertsStuartEtAl12,PinskiSimpsonStuartEtAl15}. The gpCN algorithm is also dimension-invariant, yet has the advantage of capturing posterior curvature information in a relatively inexpensive manner. It uses an approximation of the gradient and the low-rank based approximation of the Hessian at the MAP point discussed in section~\ref{sec: Laplace approximation}, rather than computing exact gradient directions and Hessian matrices at numerous samples in parameter space. The gpCN proposed sample $\pmb\eta_\text{prop}^{(k)}$ in the ${k}^\text{th}$ Metropolis Hastings (MH) iteration (described in Algorithm \ref{alg: MH}) is given by 
\begin{align}
\pmb\eta_\text{prop}^{(k)} & = \mathbf{m}_\text{MAP} + \sqrt{1-\beta^2}(\pmb\eta_\text{post}^{(k-1)}-  \mathbf{m}_\text{MAP})+\beta \pmb\eta_0^{(k)} \label{equ: gpCN Proposal}
\end{align}
where  $\pmb\eta_0^{(k)} \sim \mathcal{N}(0, \mathbf{\Gamma}_\text{Laplace,r} )$  and $\pmb\eta_\text{post}^{(k-1)}$ is the ${(k-1)}^\text{th}$ MH iteration posterior sample. The proposed sample $\pmb\eta_\text{prop}^{(k)}$ is then subject to acceptance or rejection based on the MH algorithm. The parameter $0 < \beta < 1$ is a step size that can be chosen to achieve an acceptable balance between exploring the parameter space and the acceptance rate of the MCMC chain. The smaller the value of $\beta$, the higher the acceptance rate but the longer the chain takes to explore the posterior distribution and generate independent samples. Larger $\beta$ values, however, allow the chain to visit farther points in  parameter space but lead to lower acceptance rates.  The cost of each MH iteration is dominated by the cost of evaluating $\pi_\text{post}( \pmb\eta_\text{prop}^{(k)} )$, which amounts to a single solution of the forward poroelasticity problem~\eqref{equ: Biot model b} for the log permeability realization $\pmb\eta_\text{prop}^{(k)}$. The gpCN method is integrated in \texttt{hIPPYlib} \cite{VillaPetraGhattas18}.

%%%%%%%%%%%%%%%%%%%%%%% MH algorithm 
\begin{algorithm}
\caption[]{MH algorithm for generalized preconditioned  Crank--Nicolson (gpCN) method}
\label{alg: MH}
\hspace*{\algorithmicindent} \textbf{Input} $\beta$, $\mathbf{\Gamma}_\text{Laplace,r}$, $\mathbf{m}_\text{MAP}$, \text{number-of-samples}
\begin{algorithmic}
\FOR  {$k= 1,2,3,..., \text{number-of-samples}$}
\STATE draw $\pmb\eta_0^{(k)} \sim  \mathcal{N}(0, \mathbf{\Gamma}_\text{Laplace,r}) $
\STATE calculate $\pmb\eta_\text{prop}^{(k)}$ using formula (\ref{equ: gpCN Proposal}) 
\STATE $\alpha_\text{gpCN}  \leftarrow \frac{\pi_\text{post}( \pmb\eta_\text{prop}^{(k)} )\pi_\text{Laplace,r}( \pmb\eta_\text{post}^{(k-1)} )}
                                {\pi_\text{post}(  \pmb\eta_\text{post}^{(k-1)}  )\pi_\text{Laplace,r}( \pmb\eta_\text{prop}^{(k)} )}$
\STATE draw scalar $s \sim \mathcal{U}([0,1])$
\IF  {$s < \min(1.0, \alpha_\text{gpCN})$}
\STATE  $\pmb\eta_\text{post}^{(k)}  \leftarrow \pmb\eta_\text{prop}^{(k)}$
\ELSE
\STATE  $\pmb\eta_\text{post}^{(k)}  \leftarrow \pmb\eta_\text{post}^{(k-1)}$
\ENDIF
\ENDFOR
\end{algorithmic}
\end{algorithm}

\section{Nevada Test Case}

\label{sec: Nevada Test Case}
\begin{figure}[htb] 
\includegraphics[width=1.\textwidth]{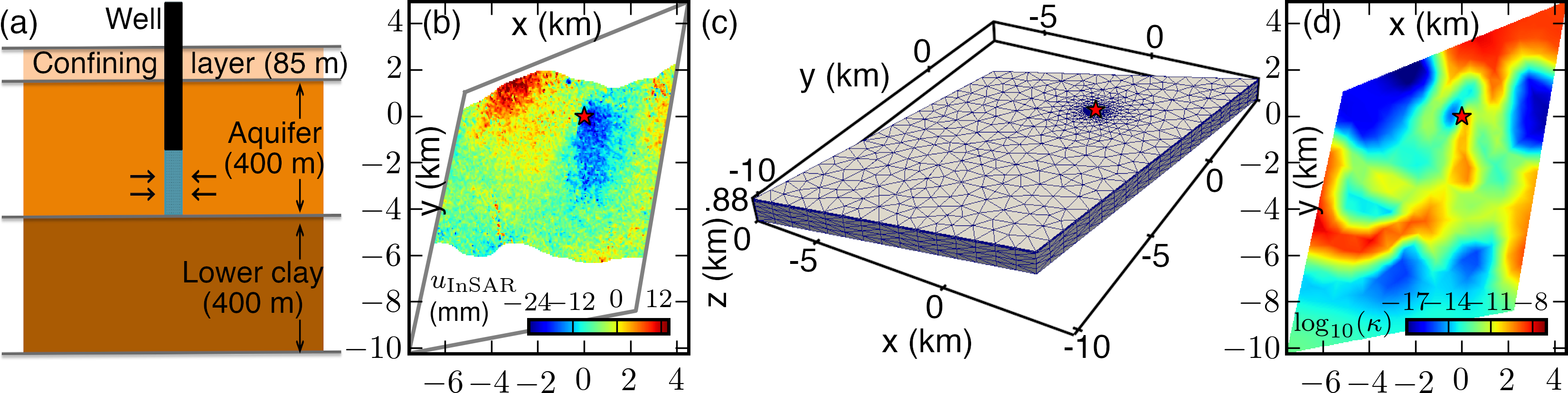}
\caption{Overview of the Nevada test case. (a) A conceptual illustration of the aquifer model. The well is screened in the lowest half of the aquifer (blue segment).
         (b) InSAR-observed LOS deformation (east, north, and vertical displacements projected onto the radar line-of-sight) between May 4, 2003 and October 26, 2003 over the study area.
         (c) The computational mesh we use to model the aquifer. The mesh has 4,081 nodes.
         (d) Two-dimensional horizontal slice of the inferred permeability, $e^{\mathbf{m}_\text{MAP}}$, at mid-aquifer depth (in decimal logarithm) using InSAR data in Figure~\ref{fig: background part I}\emph{b} and the mesh in Figure~\ref{fig: background part I}\emph{c}.  In panels (b)--(d), $x$ is the east direction and $y$ is the north direction relative to the well location $(x,y)=(0,0)$  (marked by the red star).}
\label{fig: background part I}
\end{figure}

\subsection{Study Site, InSAR Deformation Data, and Aquifer Computational Model}
\label{sec: Study Site, InSAR Deformation Data, and Aquifer Computational Model}

Our study site is located southeast of Mesquite, Nevada. The aquifer is $\sim$ 400 m thick, which is confined by an $\sim$ 85 m thick brittle layer at the top and a clay layer that is at least 400 m thick at the bottom (Figure~\ref{fig: background part I}\emph{a}). A controlled aquifer test was performed at a newly installed municipal well WX31 between May 7, 2003 and July 9, 2003 \cite{Burbey06}. This well pumped 9,028 m\textsuperscript{3} of GW per day during the study period, and it was relatively isolated from other wells in the area. The regional GW recharge is due mainly to winter precipitation in the surrounding mountains, which was negligible during the summer pumping test. \citeA{Burbey06} analyzed the aquifer geomechanical response to the pumping test using InSAR and GPS data. They found significant deviations between the observed deformation pattern and the expected deformation pattern from a homogeneous aquifer. This finding motivates us to investigate the lateral permeability heterogeneity of the aquifer using the Bayesian framework. 
 
We processed an Envisat interferogram (05/04/2003-10/26/2003) that captures the surface deformation associated with the pumping test along the radar LOS direction (Figure~\ref{fig: background part I}\emph{b}). Here the LOS unit look vector is $\alpha_\text{LOS} =[0.381, -0.08, 0.921]$ (refer to section~\ref{sec: Bayesian Framework Background} for definition), which means that the measured LOS deformation $\mathbf{d}^\text{obs}$ is mostly sensitive to vertical motion ($\alpha_3\approx 1$). The observed deformation pattern (Figure~\ref{fig: background part I}\emph{b}) shows a subsidence bowl that is  shifted toward the southeast of the well and an uplift signal in the northwest of the domain. The latter is likely due to a fault existence northwest the well \cite{Burbey08,AlghamdiHesseChenEtAl2020A}. We found that the noise in the Envisat interferogram is mainly due to phase decorrelation that is uncorrelated in space ($\sigma_\text{InSAR}=3.2$ mm). Thus we chose the noise covariance operator $\mathbf{\Gamma}_\text{noise}$ to be a diagonal matrix with diagonal values $\sigma_\text{InSAR}^2$. 

Our aquifer model is described in detail in \citeA{AlghamdiHesseChenEtAl2020A}, we assume a 3D poroelastic medium governed by the Biot model~\eqref{equ: Biot model b} and represent the pumping well as a volumetric sink term $f_p$. The parameter values of the forward model \eqref{equ: Biot model b} are provided in Table~\ref{tab: Biot Model Parameters} and  based on previously estimated values by \citeA{Burbey06}.  We discretized the model (Figure~\ref{fig: background part I}\emph{a}) using a 4,081-node unstructured tetrahedral layered mesh (Figure~\ref{fig: background part I}\emph{c}). The discretization on this mesh results in a total of ${n_\text{state}=72,980}$ pressure, displacement, and fluid flux DOFs per time step. We additionally created finer meshes of 16,896 and 67,133 nodes, which result in  ${n_\text{state}=320,824}$ and ${n_\text{state}=1,322,067}$, respectively, for the purpose of studying the eigenvalues decay of the prior-preconditioned data misfit Hessian $\mathbf{\Gamma}_\text{prior}\mathbf{H}_\text{misfit}$.
 
We imposed zero normal displacement at the bottom boundary, zero displacement at the lateral boundaries, no traction at the top boundary, and no GW flux at all boundaries. Under the assumption of zero initial deviation from the hydrostatic pressure everywhere in the domain at $t=0$, we ran the model for $T=175$ days using 154 variable-length time steps. Initial time steps are 1.2 hours to capture rapid pressure decay that occurs after the onset of pumping and increased gradually to reach five days toward the end of the simulation. Each time step involves solving a linear system of size $n_\text{state}$. %, which we assembled using the \texttt{Python}-based library \texttt{FEniCS} \cite{LoggMardalWells12}. 

\subsection{Estimating the Permeability MAP Point using the Bayesian Framework}
\label{sec: Overview of the Bayesian Framework Setup and Estimating the MAP Point (Part I)}

Based on a heuristic approach \cite{AlghamdiHesseChenEtAl2020A}, we chose the weakest bilaplacian-like Mat\'ern class prior that yet sufficiently regularizes the inverse problem to allow the (Gauss--)Newton method to converge (defined as reducing the gradient norm by four orders of magnitude). The average-over-domain pointwise standard deviation (SD) of this prior is $\sim 1.5$ in decimal logarithm, and its mean is the constant permeability value estimated by \citeA{Burbey08} as shown in Table~\ref{tab: Biot Model Parameters}. The spatial correlation length of the permeability values in these samples is $\rho = 2$ km in the lateral direction and an order of magnitude larger in the vertical direction. The latter constraint inhibits permeability variations in the vertical direction. We impose this constraint because (1) vertical variations are likely not informed by deformation data obtained on the surface; (2) the large aspect ratio of the aquifer horizontal-to-vertical dimensions leads to predominantly horizontal flow.

In the Part I paper \cite{AlghamdiHesseChenEtAl2020A}, we used the InSAR surface deformation data (Figure~\ref{fig: background part I}\emph{b}) to solve for the permeability realization that is most likely to be the true permeability field. This is known as the maximum a posteriori (MAP) point of the posterior distribution (Equation~\eqref{equ: Bayes Formula Finite}). We found that the MAP point (Figure~\ref{fig: background part I}\emph{d}) reveals distinct features in the lateral permeability field: high permeability channel extending from south the well to the southwest and low permeability barrier to the northwest of the well. We validated the InSAR-based permeability MAP point solution using an independent GPS data set as well as a synthesized data set. We computed the MAP point using (Gauss--)Newton method and observed a mesh-independent convergence. We also demonstrated the consistency of the solution with respect to a wide range of prior assumptions, and InSAR data multi-look choices, given a consistent data noise treatment.

\subsection{Creating Low-Rank Based Laplace Approximation of the Posterior and MCMC Sampling}
\label{sec: Sampling and Postprocessing (Part II)}
The Bayesian inversion framework allows us to quantify the uncertainty in the solution by sampling the posterior distribution. To estimate the uncertainty in the aquifer permeability, we first build the low-rank based Laplace approximation $\pi_\text{Laplace,r}$ of the posterior, a process that entails solving the generalized eigenvalue problem \eqref{equ: Generalized Eigenvalue Problem} using a randomized generalized eigensolver \cite{VillaPetraGhattas2020}, to obtain the low-rank based approximation of posterior covariance matrix \eqref{equ: Low Rank Based Lapalce Approximation}. We study two approximations in which we use truncation values $r=80$ and $r=150$, respectively. This low-rank based Laplace approximation is feasible to compute---it mainly costs $O(r)$ incremental forward \eqref{equ: Incremental Forward} and incremental adjoint  \eqref{equ: Incremental Adjoint} solves---and can serve (depending on how close the parameter to observable map is to a linear one) as a good approximation of the posterior distribution $\pi_\text{post}$. Additionally, we use this approximation in the gpCN proposal (see Algorithm~\ref{alg: MH}) to sample the true posterior.

To that end, we generate five MCMC chains, using the gpCN MCMC method (Algorithm~\ref{alg: MH}), each of length 50,000,  with additional 1,000 burn-in samples. Each chain start point $\pmb\eta_\text{post}^{(0)}$ is an independent sample from the low-rank based Laplace approximation $\pi_\text{Laplace,r}$, with $r=80$. We similarly generate two pCN chains \cite{CotterRobertsStuartEtAl12}. Generating a single gpCN chain (and similarly a pCN chain) costs 50,000 forward solves of the system~\eqref{equ: Biot model b}, discretized on the 4,081-node mesh. These solves are required to evaluate the ratio $\alpha_\text{gpCN}$ in  Algorithm~\ref{alg: MH}. We store pressure $p$ and displacement $\mathbf{u}$ solutions evaluated at time $t_\text{InSAR}\approx5.7$ months, the time at which the LOS deformation (Figure~\ref{fig: background part I}\emph{b}) is observed relative to the start of the simulation, for different realizations of $\mathbf{m}$. We set the step size $\beta=0.25$ in gpCN proposal formula~\eqref{equ: gpCN Proposal} and set $\beta=0.01$ in pCN proposal. These values were found with a trial-and-error approach to balance between high acceptance rate and large step size. To compare several posterior-based QoI vs prior-based QoI,  we generate 1,000 samples from the prior distribution and the low-rank based Laplace approximation (with $r=80$) directly and solve the system~\eqref{equ: Biot model b} at each sample. We carry out the implementation using the \texttt{FEniCS} library for finite element discretization in space \cite{LoggMardalWells12} and the \texttt{hIPPYlib} library for state-of-the-art Bayesian and deterministic PDE-constrained inversion algorithms \cite{VillaPetraGhattas18}.

%%%%%%%%%%%%%%%%%%%%%%%%%%%%%%%%%%%%%%%%%%%%%%%%%%%%%%%%%%%%%%%%%%%%%%%%%%%%%%%
%%%%%%%%%%%%%%%%%%%%% Section: Results %%%%%%%%%%%%%%%%%%%%%%%%%%%%%%%%%%%%%%%%
%%%%%%%%%%%%%%%%%%%%%%%%%%%%%%%%%%%%%%%%%%%%%%%%%%%%%%%%%%%%%%%%%%%%%%%%%%%%%%%
\section{Results and Discussion}
\label{sec: Results and Discussion}

Here we first establish the low rank property of the data misfit Hessian for the Nevada test-case poroelasticity problem introduced in section~\ref{sec: The Prior Preconditioned Data Misfit Hessian Spectrum}. This property is critical for the applicability of the gpCN algorithm.  Then we study the convergence and mixing of the gpCN algorithm and compare it to the performance of the pCN algorithm in section~\ref{sec: gpCN Performance}. Finally, we use the gpCN algorithm to quantify the uncertainty in the aquifer permeability and state variables QoI, given the InSAR data information, in sections~\ref{sec: Uncertainty Quantification of the Aquifer Permeability} and \ref{sec: Uncertainty Quantification of the State Variables}.

%******************************************************************************
% The Prior Preconditioned Data Misfit Hessian Spectrum
%******************************************************************************
\subsection{ The Prior Preconditioned Data Misfit Hessian Spectrum}
\label{sec: The Prior Preconditioned Data Misfit Hessian Spectrum}

\begin{figure}[htb]
\includegraphics[width=1.\linewidth]{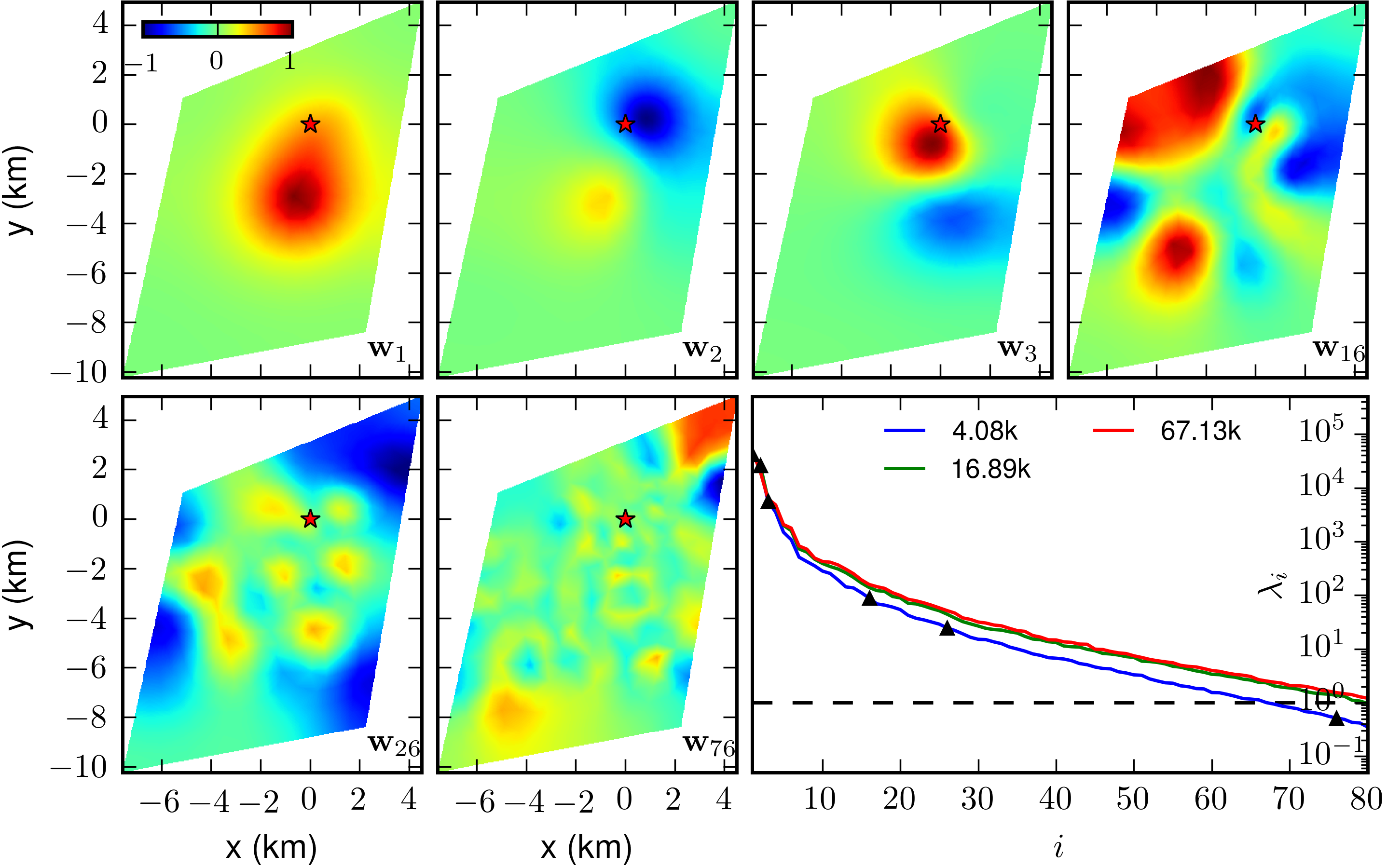}%
\caption{Eigenvectors $\mathbf{w}_i$ for ${i=1,2,3,16,26,76}$ of the generalized eigenproblem \eqref{equ: Generalized Eigenvalue Problem} using InSAR data and the $4{,}081$-node mesh. (Bottom right panel) Eigenvalues of the data misfit Hessian using the 4,081-node (blue), the 16,896-node (green), and the 67,133-node (red) meshes. The black dashed line is ${\lambda_i=1}$. The eigenvalues corresponding to the eigenvectors $\mathbf{w}_i$ for ${i=1,2,3,16,26,76}$ in the $4{,}081$-node mesh are marked by black triangles. In all meshes, only $\sim$70--80 eigendirections can be inferred from the data (i.e.\ those inferred with high confidence, corresponding to $\lambda_i> 1$).}
\label{fig: Eignvalues Decay and Eigenvectors}
\end{figure}

The plausibility of the low-rank based Laplace approximation and the computational feasibility of the gpCN method, discussed in sections \ref{sec: Laplace approximation} and \ref{sec: gpCN}, rely on the inherent low rank structure of the data misfit Hessian. This property allows for truncating the low-rank based Laplace approximation at $r\ll n_n$. In section~\ref{sec: Laplace approximation} we proved that this property holds for an idealized 1D poroelasticity problem. To demonstrate this property for the 3D Nevada test problem introduced in section~\ref{sec: Nevada Test Case} we computed the eigenvalues and eigendirections of the prior-preconditioned data misfit Hessian evaluated at the MAP point (Figure~\ref{fig: Eignvalues Decay and Eigenvectors}). The bottom right panel shows the decay of the eigenvalues on increasingly refined meshes. The decay patterns in the 16,896-node and 67,133-node meshes are almost identical, which demonstrates a mesh-independent eigenvalue decay on sufficiently resolved meshes. On all three meshes, only about $r=70$--$80 \ll n_n$ eigenvalues are larger than one, and therefore, only the $r$ eigendirections that correspond to these eigenvalues can be inferred from InSAR data with high confidence. Selected eigendirections are shown in Figure~\ref{fig: Eignvalues Decay and Eigenvectors}.  Similar to the 1D problem in section \ref{sec: Laplace approximation}, the larger the eigenvalue the smoother the corresponding eigendirection, indicating that it is easier to infer smooth modes. The distinctive features of the first few eigendirections, which are inferred strongly from the data, are localized near the well and the surrounding subsidence bowl, where surface deformation is largest.   

%******************************************************************************
% gpCN Performance
%******************************************************************************
\subsection{Performance of the gpCN algorithm}
\label{sec: gpCN Performance}

\begin{figure}[htbp]
\centering
\includegraphics[width=1\linewidth]{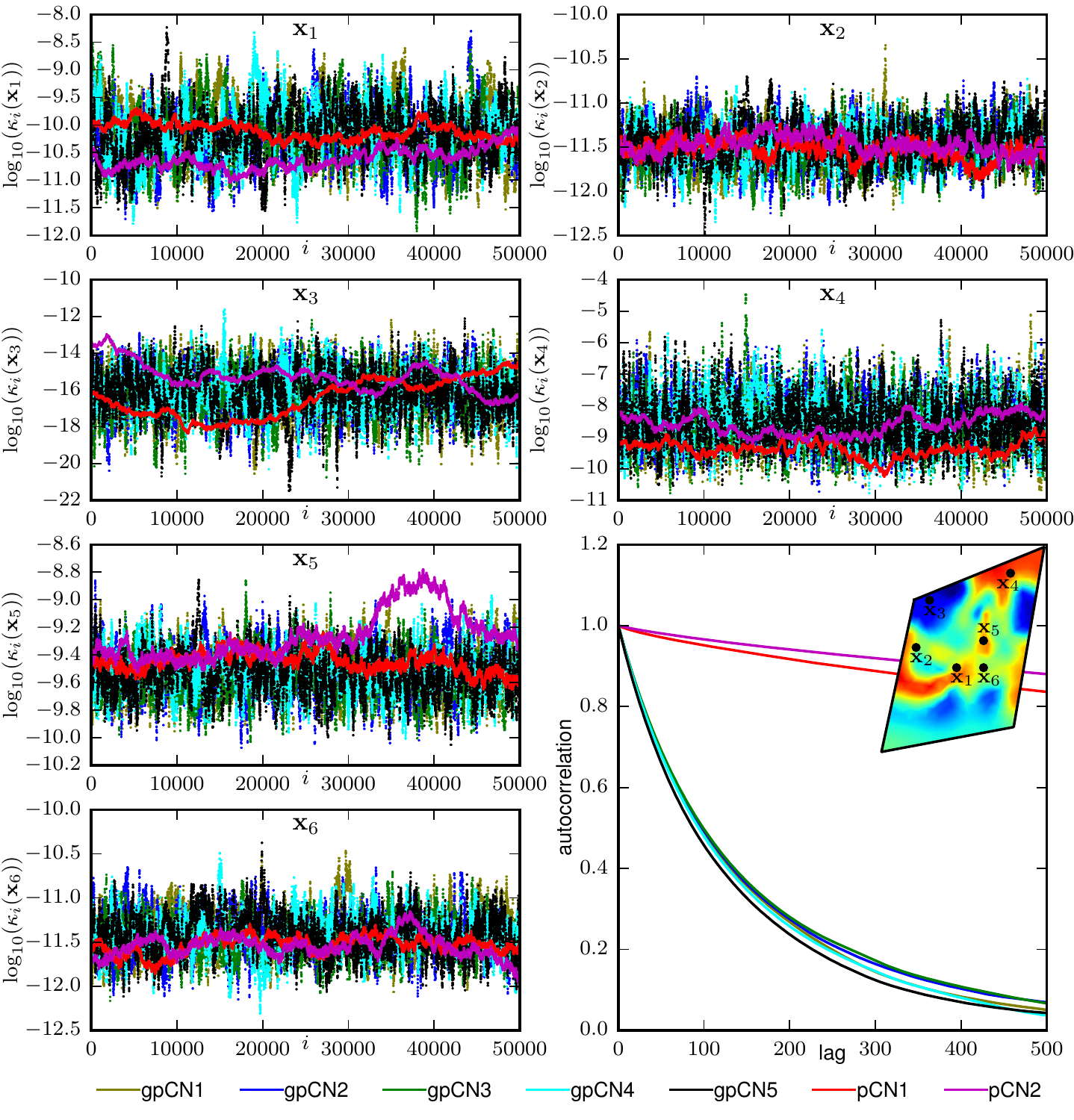}
\caption { pCN and gpCN chains mixing at selected points in space $\mathbf{x}_1=(-2.0,-4.0,0.6)$, $\mathbf{x}_2=(-5.0,-2.5,0.6)$, $\mathbf{x}_3=(-4.0,1.0,0.6)$, $\mathbf{x}_4=(2.0,3.0,0.6)$, $\mathbf{x}_5=(0,-2.0,0.6)$, and $\mathbf{x}_6=(0,-4.0,0.6)$ in km. (Bottom right panel) autocorrelation versus lag. Autocorrelation values are reported as average over the domain. The two pCN chains and the five gpCN chains are labeled pCN1, pCN2, and  gpCN1--gpCN5, respectively.} 
\label{fig: Parameter Mixing}
\end{figure}

For uncertainty quantification in realistic 3D transient multi-physics problems it is essential to improve the sampling of the parameter space.  Incorporating log posterior Hessian information in gpCN proposal, through the low-rank based Laplace approximation, leads to better convergence properties, compared to the pCN algorithm, see Table~\ref{tab: chains convergence} for methods performance. The gpCN algorithm achieves an acceptance rate of  $\sim 22\%$ when using a step size $\beta=0.25$. To achieve a similar acceptance rate for the pCN algorithm, $\sim 24\%$, a much smaller number $\beta=0.01$ is required, which leads to smaller MCMC steps and slower exploration of the feasible parameter space. Consequently, the sample autocorrelation in gpCN chains decays much faster than in pCN chains, as shown in the bottom right panel in Figure~\ref{fig: Parameter Mixing}. The large step-size in gpCN chains increases the number of independent MCMC samples and leads to a larger effective sample size compared to pCN chains (Table~\ref{tab: chains convergence}). Visualizing chains mixing at selected points in space (Figure~\ref{fig: Parameter Mixing}) highlights the performance difference between the two methods. The pCN chains vary slowly, fail to converge and do not explore the feasible parameter space while gpCN chains move more rapidly and exhibit more of a ``fuzzy worm" pattern  \cite{Bui-Thanh12} in which  correlation between samples is weaker.

\begin{table}[tbh]
\caption{pCN and gpCN performance} 
\begin{center}
\begin{tabular}{ l|c|c|c|c|c|c }
\hline
 Method &  \# chains$^{a}$  & PDE solves per chain & $\beta$ & Acceptance rate & IAT$^{b}$ & ESS/chain$^{c}$    \\
\hline
 pCN       &  2          & 50,000           & $0.01$ &  $\sim 24\%$    & 9754 & 5    \\
 gpCN      &  5          & 50,000           & $0.25$ &  $\sim 22\%$    & 494 & 101    \\
\hline
\multicolumn{7}{p{13cm}} {$^{a}$ Each chain is of length 50,000 samples.} \\
\multicolumn{7}{p{13cm}} {$^{b}$ IAT is the pointwise integrated autocorrelation time reported as average over the domain and over the number of chains.} \\
\multicolumn{7}{p{13cm}} {$^{c}$ $\text{ESS}=\frac{\text{Chain length}}{\text{IAT}}$ is the effective sample size.} \\
\end{tabular}
\end{center}
\label{tab: chains convergence}
\end{table}

%******************************************************************************
% Uncertainty Quantification of the Aquifer Permeability
%******************************************************************************
\subsection{Uncertainty Quantification of the Aquifer Permeability}
\label{sec: Uncertainty Quantification of the Aquifer Permeability}

In this section we present the solution of the Bayesian inverse problem through the variance, samples, and selected marginal distributions of the permeability posterior distribution. In particular we study both the low-rank based Laplace approximation and the MCMC-based posterior distributions. 

\subsubsection{Permeability Pointwise Standard Deviation and Samples}
\label{sec: Permeability Pointwise Standard Deviation and Samples}

\begin{figure}[htbp]
\centering
\includegraphics[width=1\linewidth]{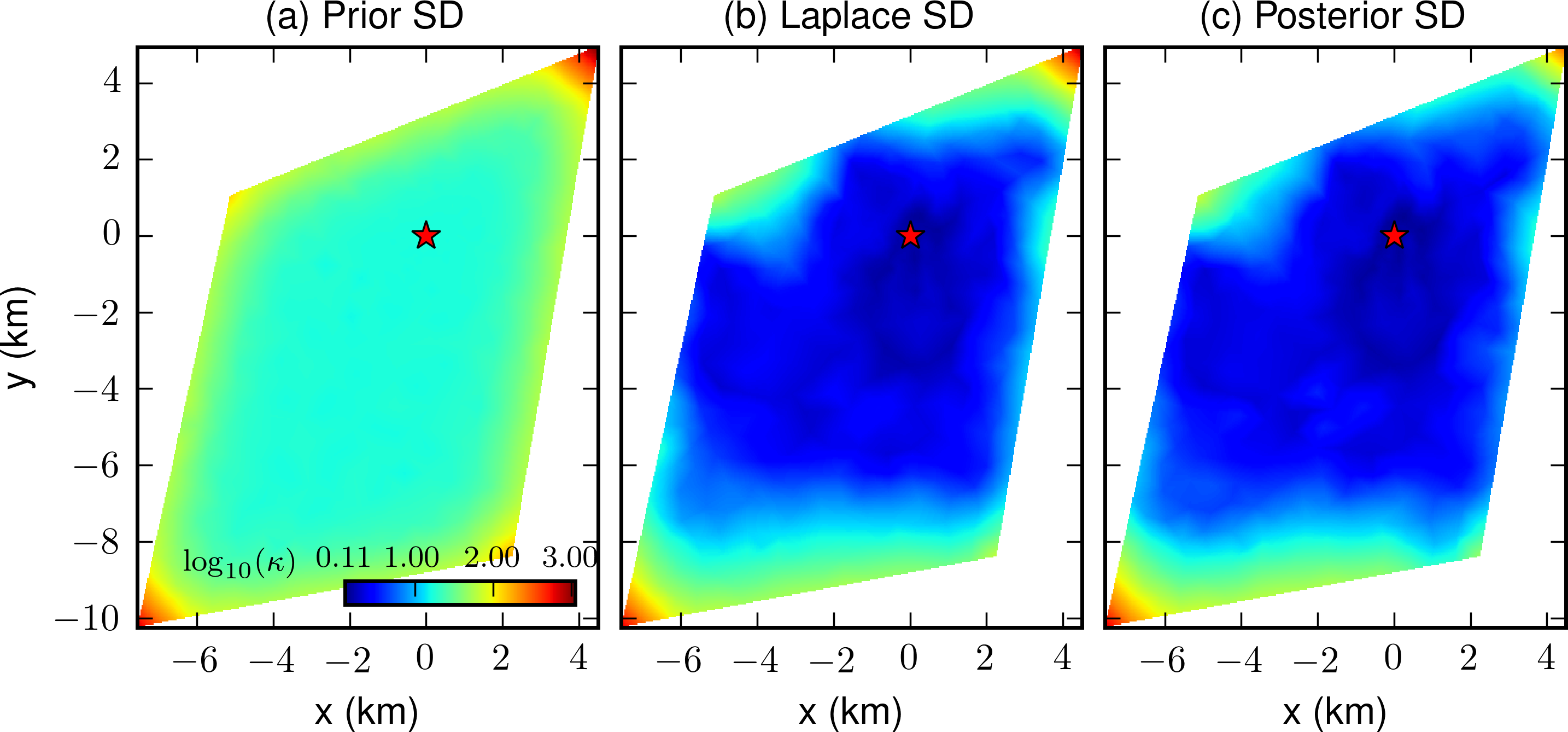}
\caption[Prior, Low-Rank Based Posterior Laplace Approximation, and True Posterior Pointwise Standard Deviation]
{Two-dimensional horizontal slices of the permeability pointwise standard deviation (in decimal logarithm) at mid-aquifer depth of (a) the prior distribution, (b) the low-rank based Laplace approximation ($r=80$), and (c) the MCMC-based true posterior. In all the panels, $x$ is the east direction and $y$ is the north direction relative to the well location $(x,y)=(0,0)$ (marked by the red star).}
\label{fig: Pointwise Variance}
\end{figure}   

\begin{figure}[htbp]
\centering
\includegraphics[width=1\linewidth]{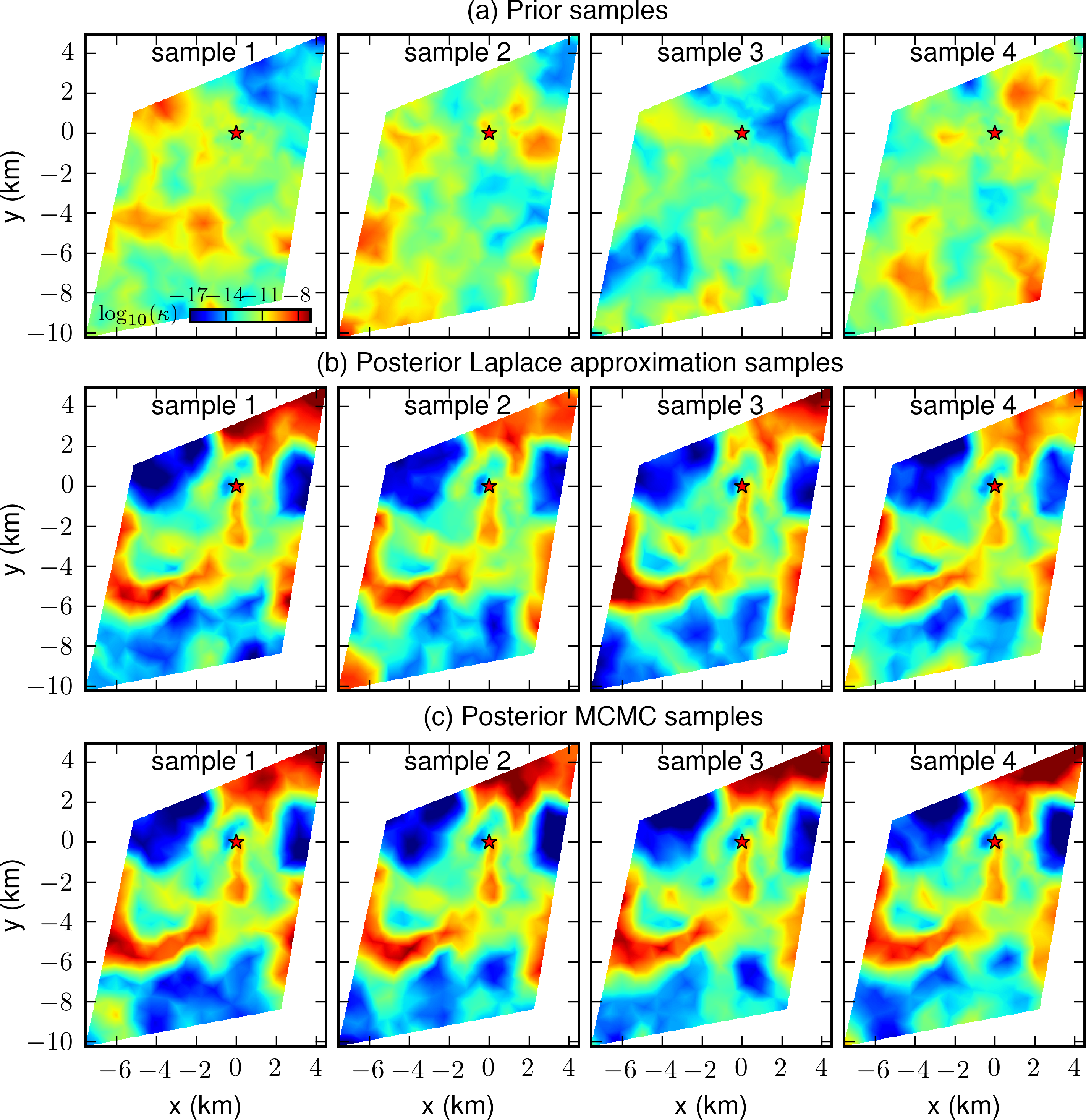}
\caption[Prior, Low-Rank Based Posterior Laplace Approximation, and True Posterior Permeability Samples]
{Two-dimensional horizontal slices of permeability samples (in decimal logarithm) at mid-aquifer depth from (a) the prior distribution, (b) the low-rank based posterior Laplace approximation ($r=80$), and (c) the true posterior distribution (MCMC samples).}
\label{fig: Samples}
\end{figure}

The posterior distribution incorporates information from InSAR data and the aquifer model. This significantly reduces the pointwise standard deviation (SD) of the permeability posterior distributions compared to the prior pointwise SD in most of the domain. Figure~\ref{fig: Pointwise Variance} shows this is the case for both the low-rank based Laplace approximation and the MCMC-based posterior. The spatial variation of the SD inferred from the low-rank based Laplace approximation  in panel~\ref{fig: Pointwise Variance}\emph{b} is very similar to the SD of the MCMC-based true posterior distribution in panel~\ref{fig: Pointwise Variance}\emph{c}. The MCMC-based true posterior, however, infers a slightly larger area where permeability can be inferred with high confidence.  This is also reflected in the samples obtained from these three distributions. Samples from the prior distribution exhibit varying random patterns (Figure~\ref{fig: Samples}\emph{a}) whereas samples from the low-rank based Laplace approximation and the MCMC-based true posterior (Figure~\ref{fig: Samples}\emph{b}~and~\ref{fig: Samples}\emph{c}) show consistent features, e.g.\ a low permeability area northwest the well and a high permeability channel extending toward the south/southwest. In the latter two distributions, permeability patterns at the southern part of the domain display more randomness than in the center and the north. This randomness is expected because the SD at the southern part is relatively larger (Figures~\ref{fig: Pointwise Variance}\emph{b}~and~\ref{fig: Pointwise Variance}\emph{c}) and is likely a result of two factors, the southern part of the domain is far from the major part of the subsidence bowl, and the InSAR data we use in the inversion does not cover that area of the domain.

\subsubsection{Location-wise Permeability PDF Marginals}
\label{sec: Location-wise Permeability PDF Marginals}
\begin{figure}[htbp]
\centering
\includegraphics[width=1\linewidth]{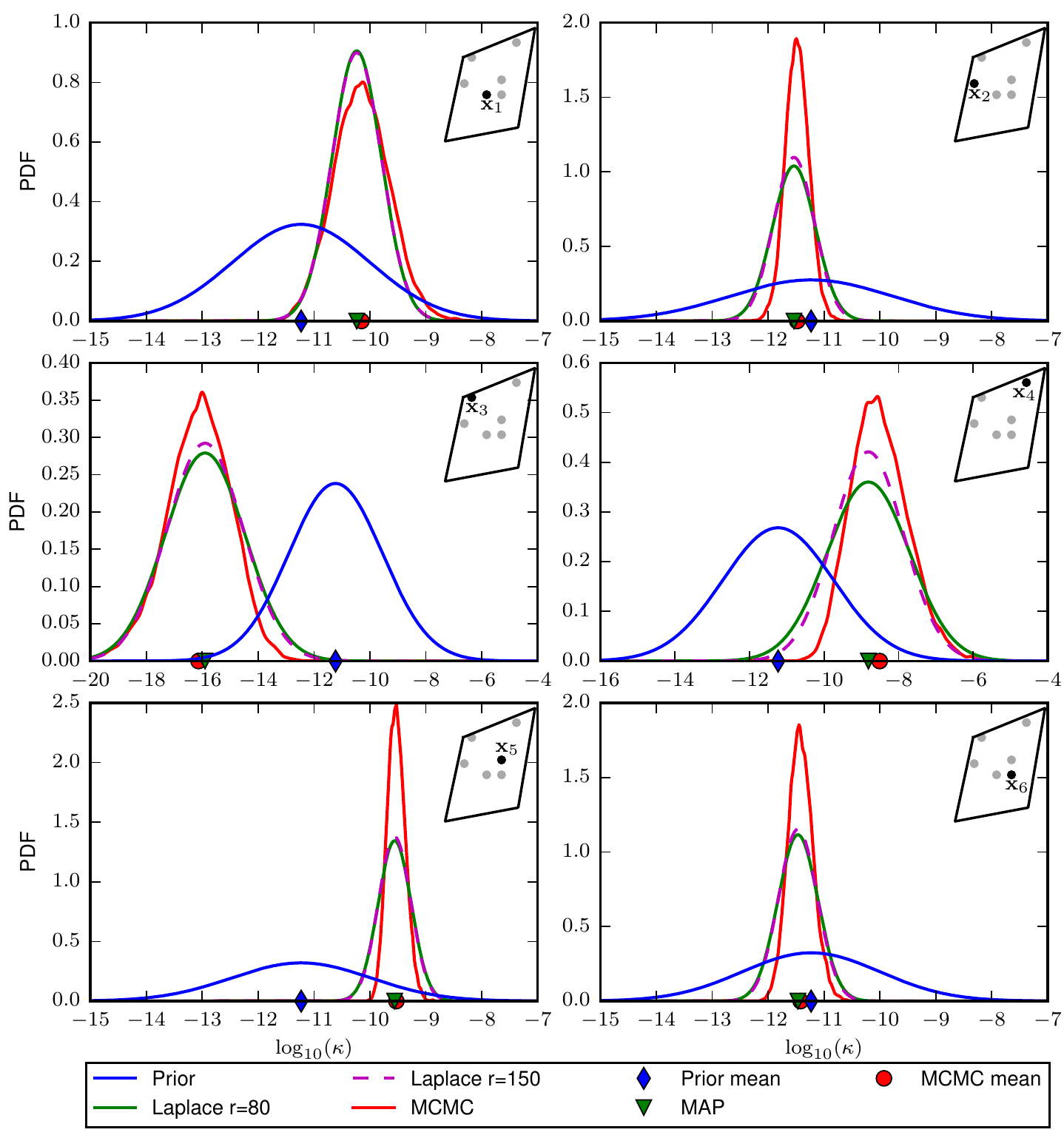}
\caption{One-dimensional marginals at selected points $\mathbf{x}_i$ for $i=1,2,...,6$ (coordinates are provided in Figure~\ref{fig: Parameter Mixing}) of the prior distribution, the low-rank based Laplace approximation (using $r=80$ and $r=150$), and the MCMC-based true posterior distribution.}
\label{fig: 1D marginals}
\end{figure}

To illustrate the broad range of results we compare 1D marginals of the prior distribution to posterior distributions inferred using both the low-rank based Laplace approximation  and MCMC at selected points in the domain ($\mathbf{x}_i$ for $i=1,2,...,6$ in Figure~\ref{fig: 1D marginals}). The posterior distributions inferred from  the Laplace approximations with $r=80$ and $r=150$ modes are very similar. This demonstrates that most of the information from the data is contained in the first 80 modes and adding the next 70 modes in the low-rank based Laplace approximation provides only limited additional information. The marginals  of the MCMC-based true posterior distribution at points $\mathbf{x}_i$ are obtained by applying 1D Gaussian kernel density estimation (KDE) to the pointwise values of the MCMC samples  at $\mathbf{x}_i$. These marginals confirm the observation from the pointwise SD values (Figure~\ref{fig: Pointwise Variance}) that the posterior distribution is significantly narrower than the prior distribution. Moreover, the pointwise marginals reveal that the MCMC-based true posterior exhibits more certainty in the permeability inference compared to the low-rank based Laplace approximation in most of the selected pointwise locations ($\mathbf{x}_2$ to $\mathbf{x}_6$).

The certainty in the inferred permeability varies with location in the domain as shown in Figure~\ref{fig: 1D marginals}. We notice, for example, inference with high certainty at the location $\mathbf{x}_5$, which is located in the high permeability channel extending south from the well and within the subsidence bowl. On the other hand, the existence of a low permeability region northwest of the well reduces the GW flux and increases the uncertainty in the parameters in the region beyond this flow barrier, for example at location $\mathbf{x}_3$. This effect was also observed by \citeA{HesseStadler14} in synthetic model problems and we refer to it here as the shadowed region. The uncertainty in the permeability of the shadowed region is large, because in the absence of flow there is no surface deformation that informs the model permeability in this region.

\begin{figure}[htbp]
\centering
\includegraphics[width=1\linewidth]{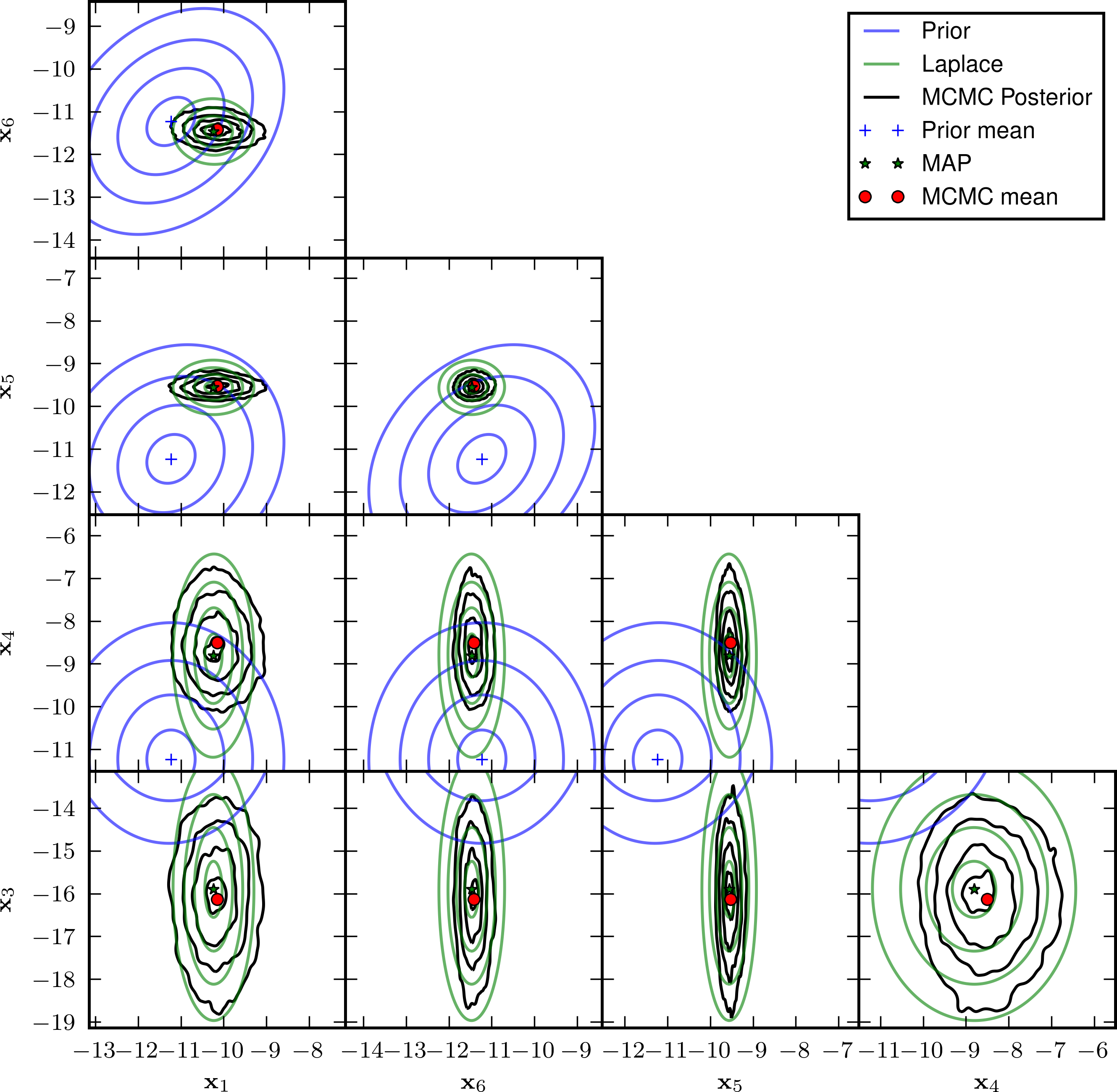}
\caption[Prior, Low-Rank Based Posterior Laplace Approximation, and True Posterior Two-Dimensional Marginals at Selected Pairs of Points in Space $\{\mathbf{x}_i,\mathbf{x}_j\}$]
{Two-dimensional marginals at selected pairs of points in space $\{\mathbf{x}_i,\mathbf{x}_j\}$ for $i,j=1,3,4,5,6$ and $i \neq j$ (coordinates are provided in Figure~\ref{fig: Parameter Mixing}) of the prior, the low-rank based Laplace approximation ($r=80$), and the MCMC-based true posterior. The values of the contour lines of each distribution  are $c_\text{cl}\times \max(\Pi) $ for $c_\text{cl} = 0.1, 0.3, 0.6, 0.9$, where $\Pi$ is the distribution PDF.}
\label{fig: 2D marginals}
\end{figure}

Slight deviation of the MCMC-based true posterior from a Gaussian distribution can be observed  from the posterior two-dimensional marginals (see Figure \ref{fig: 2D marginals}). The mean of the marginals at $\{\mathbf{x}_4,\mathbf{x}_1\}$ and $\{\mathbf{x}_3,\mathbf{x}_4\}$ is different from the MAP point and the distribution  displays right-skewness of the marginal at $\{\mathbf{x}_3,\mathbf{x}_4\}$.  On the whole though, the deviations of the MCMC-based true posterior from Gaussian distribution are minor in this particular case.

\begin{figure}[htbp]
\centering
 \includegraphics[width=1.\textwidth,draft=false]{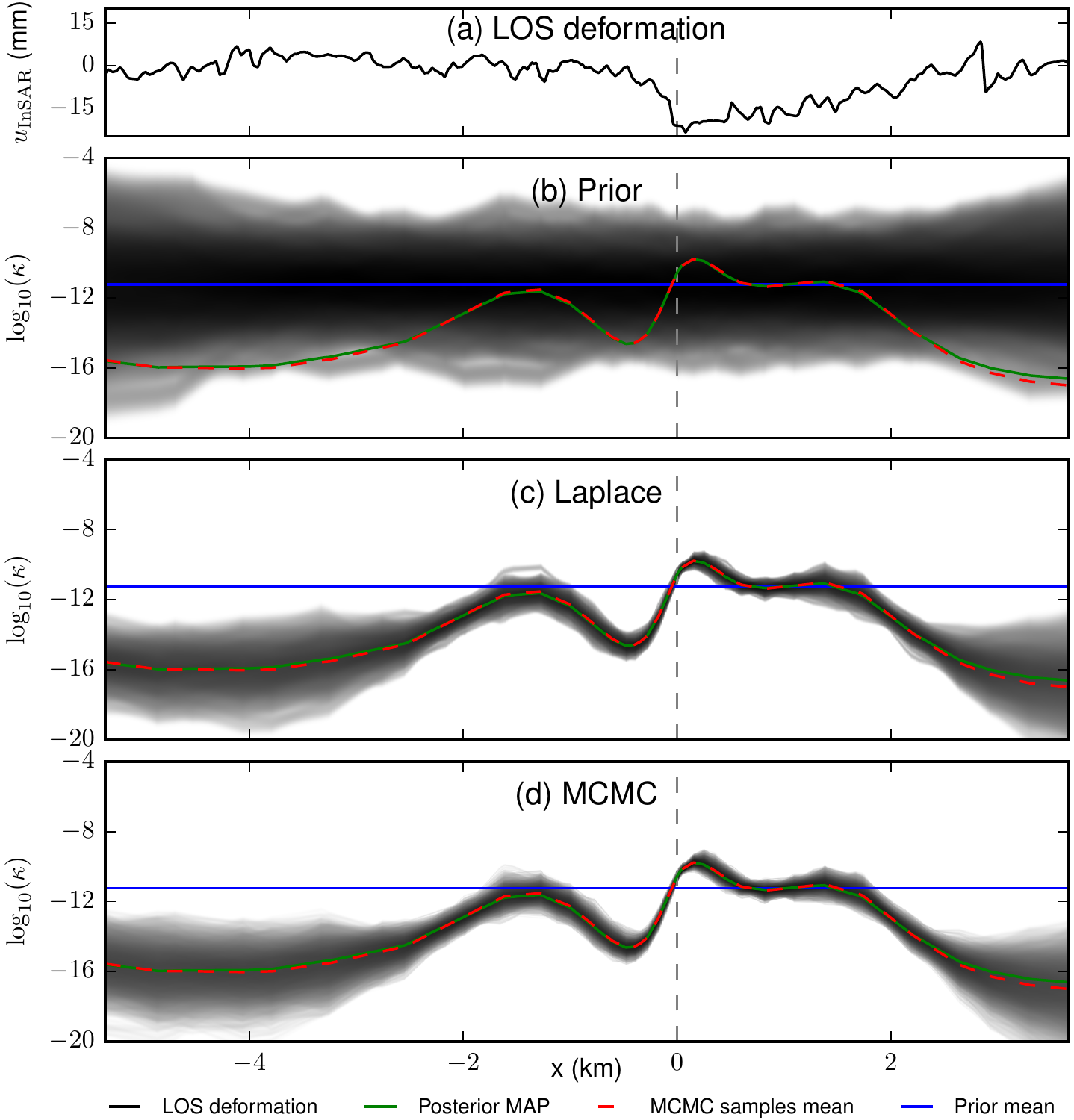}
\caption{Prior, low-rank based Laplace approximation, and true posterior marginals over a line. (a) InSAR LOS deformation signal over a line segment $l_1$ parallel to the $x$ axes, extends from the west to the east of the domain, and passes by the well location at the surface. Marginal distributions, over a line segment $l_2$ parallel to the $x$ axes, extends from the west to the east of the domain, and  passes by the well location at $z=.6$ km, computed from (b) the prior, (c) the low-rank based Laplace approximation, and (d) the MCMC-based true posterior. We apply KDE to pointwise values of the samples at evenly spaced 1,000 grid points of $l_2$ to generate the approximate  marginal PDFs. The gray shade is proportional to the PDF value. Values of the estimated PDF beyond the shaded area are negligible. The well location is marked with a gray dotted vertical line and the prior mean, the MAP point, and the MCMC samples mean are shown in blue, green and dotted red, respectively (in decimal logarithm).}
\label{fig: Marginal Over Line}
\end{figure}

Figure~\ref{fig: Marginal Over Line} shows marginals over a line segment parallel to the $x$ axes that extends from the west to the east of the domain and passes by the well at point $(x,y,z) = (0,0,.6 \text{ km})$.    We observe that, compared to the prior distribution (Figure~\ref{fig: Marginal Over Line}\emph{b}), the posterior distributions characterize the permeability with much more certainty especially in the middle part of the line segment near the well from $\sim -2$ to $\sim 2$ km (Figures~\ref{fig: Marginal Over Line}\emph{c}~and~\ref{fig: Marginal Over Line}\emph{d}). This is mainly because this interval coincides with  significant deformation signal or deformation gradient (Figure~\ref{fig: Marginal Over Line}\emph{a}). In particular, the sharp drop in the permeability just west to the well location is inferred with very high certainty as evident by the narrow marginal distribution at that location. The  location of this feature coincides with large gradient in the subsidence from almost no deformation just west to the well to significant subsidence at and east to the well. This demonstrates that large gradients in the deformation data inform the permeability characterization strongly.

\begin{figure}[htbp]
\centering
\includegraphics[width=1\linewidth]{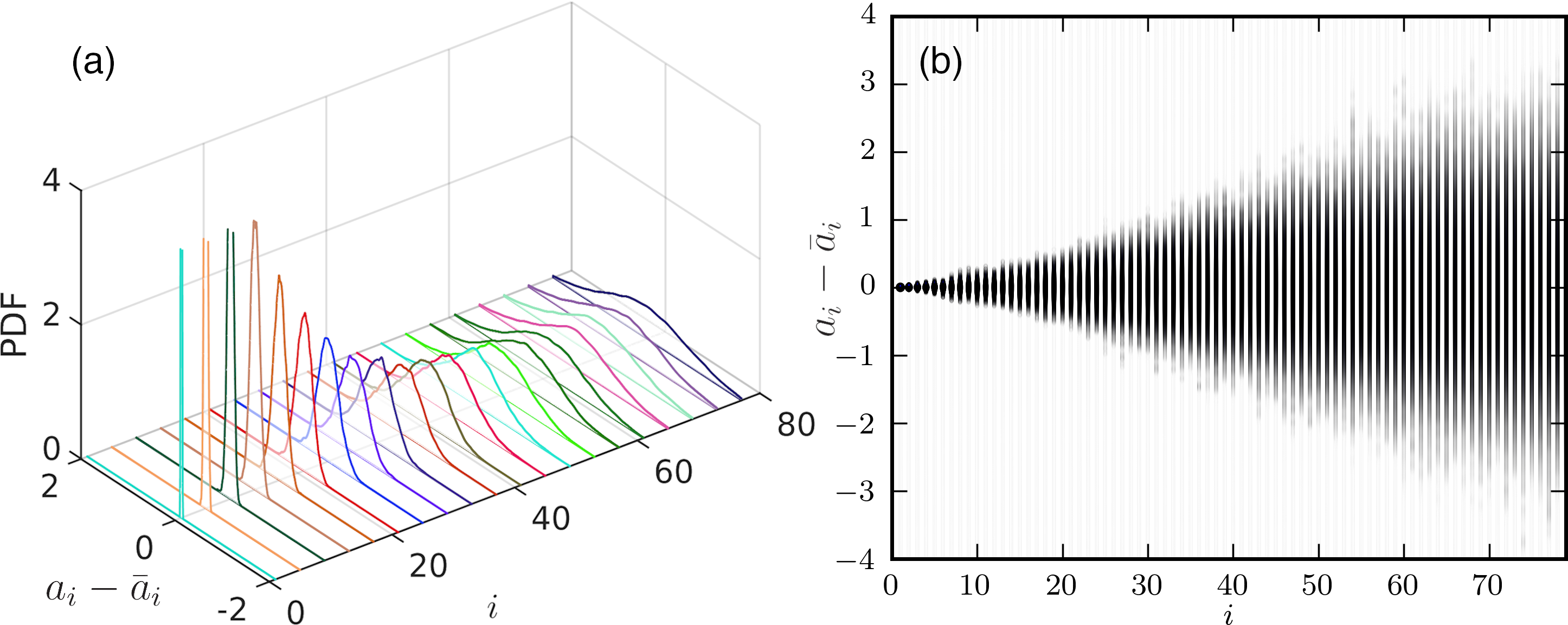}
\caption{Marginal distributions of the gpCN MCMC samples components $a_i$ in the first $r=80$ eigendirections $\mathbf{w}_i$, for $i=1,2,...,r$. The PDFs are shifted by their means $\bar{a}_i$. (a) PDFs of  $a_i$ (for better visualization, we display only every fourth PDF and truncate the PDF plot at $\text{PDF}=4$). (b) PDFs of  $a_i$ are plotted vertically in gray scale. The darker the gray shade, the higher the PDF value. Values of the estimated PDF beyond the shaded area are negligible (less than $\sim .0003$).}
\label{fig: Components Marginals}
\end{figure}

\subsubsection{Permeability PDF Marginals Over the Eigendirections $\mathbf{w}_i$}
\label{sec: Location-wise Permeability PDF Marginals}

In the preceding discussion, we studied the permeability characterization location-wise in the physical domain. Uncertainty in the permeability characterization can be viewed from a different perspective, in the directions of the parameter space that are determined by the generalized eigenvalue problem~\eqref{equ: Generalized Eigenvalue Problem} \cite{PetraMartinStadlerEtAl14}. To that end, we compute the marginals of the MCMC-based posterior distribution in the first $r=80$ eigendirections $\mathbf{w}_i$ (we show selected directions in Figure~\ref{fig: Eignvalues Decay and Eigenvectors}). We denote with $a_i$, for $i=1,2,...,r$, the components of the $j^\text{th}$ MCMC sample $\pmb\eta_\text{post}^{(j)}$ in the eigendirections $\mathbf{w}_i$, $j=0,1,...,n_s$ where $n_s = 250,000$ is the total number of gpCN MCMC samples. We can write the sample $\pmb\eta_\text{post}^{(j)}$ as a linear combination of the eigendirections $\mathbf{w}_i$ as follows: 
\begin{align}
\pmb\eta_\text{post}^{(j)} = \sum_{i=1}^{n_n} a_i \mathbf{w}_i %+  a_1 \mathbf{w}_1   + ... +  a_{n_n}\mathbf{w}_{n_n}.
\end{align}
Using the $\mathbf{\Gamma}_\text{prior}^{-1}$-orthogonality of the directions $\mathbf{w}_i$, the components $a_i$ are given by $a_i =  \mathbf{w}_i^T\mathbf{\Gamma}_\text{prior}^{-1}\pmb\eta_\text{post}^{(j)}$. We use 1D Gaussian KDE to approximate the distribution of each component $a_i$ over the samples $\pmb\eta_\text{post}^{(j)}$ for $j=0,1,...,n_s$. We show $a_i$ distributions  in Figure~\ref{fig: Components Marginals}, shifted by their averages $\bar{a}_i$ for ease of comparison. We notice that, as expected, the more dominant the eigendirection in the generalized eigenvalue problem~\eqref{equ: Generalized Eigenvalue Problem}, the more certain the inference of the component in that direction. The first few directions, in particular, are inferred with significantly high confidence. We also observe that the distributions of some $a_i$'s deviate slightly from a Gaussian distribution showing a mild skewness.

%******************************************************************************
% Uncertainty Quantification of the State Variables
%******************************************************************************
\subsection{Uncertainty Quantification of the State Variables}
\label{sec: Uncertainty Quantification of the State Variables}

\begin{figure}[htbp]
\centering
\includegraphics[width=1\linewidth,draft=false]{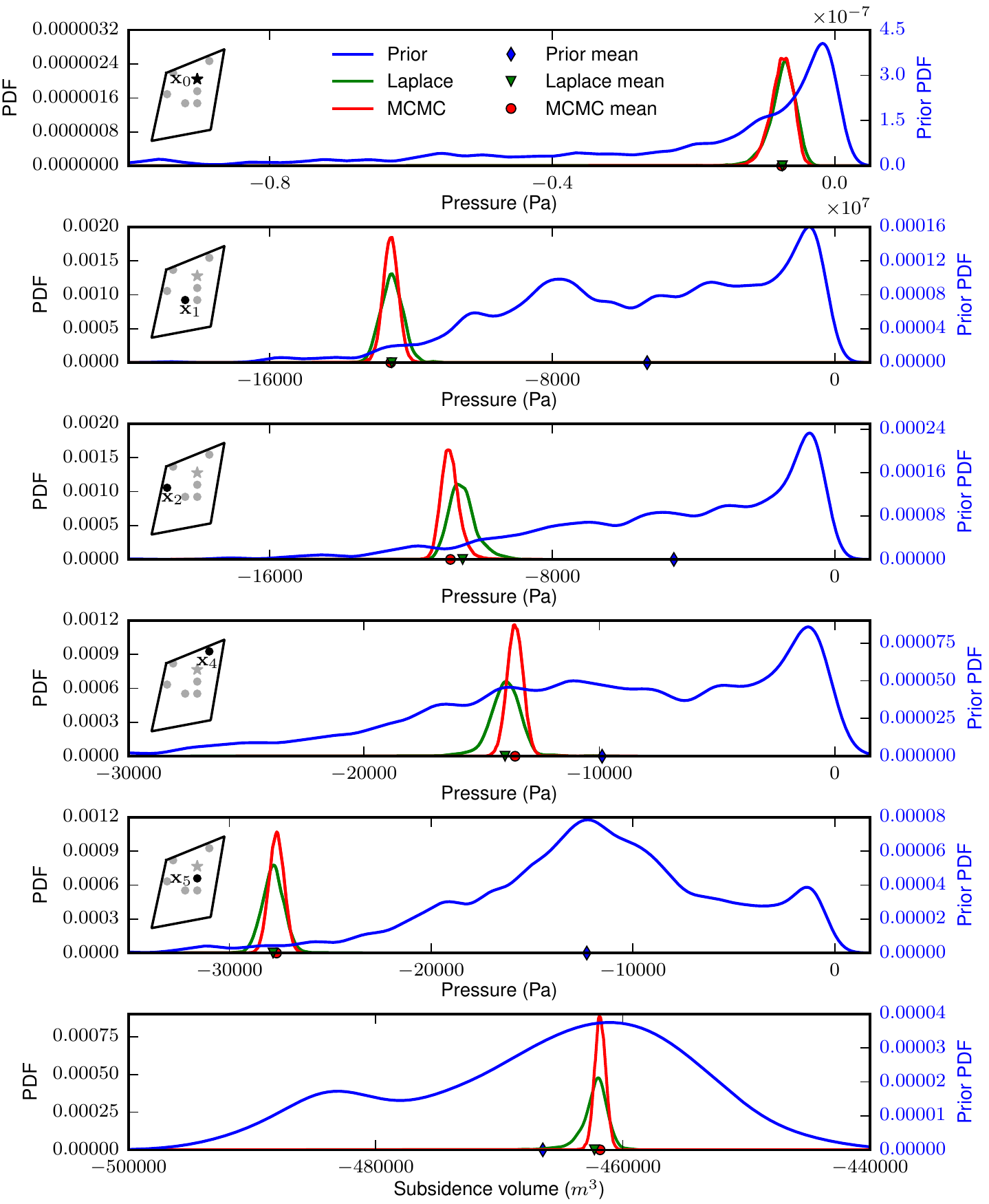}
\caption
{Prior, low-rank based Laplace approximation, and MCMC-based posterior predictives for pointwise pressure values at selected locations ($\mathbf{x}_0, \mathbf{x}_1, \mathbf{x}_2, \mathbf{x}_4$, and $\mathbf{x}_5$) and total subsidence volume (bottom panel). In all panels, the values of the Laplace approximation-based and MCMC-based PDFs are measured by the left $y$-axis scale while the right $y$-axis scale measures the prior PDF.}
\label{fig: QoI Marginals}
\end{figure}

\begin{figure}[htbp]
\centering
\includegraphics[width=1\linewidth,draft=false]{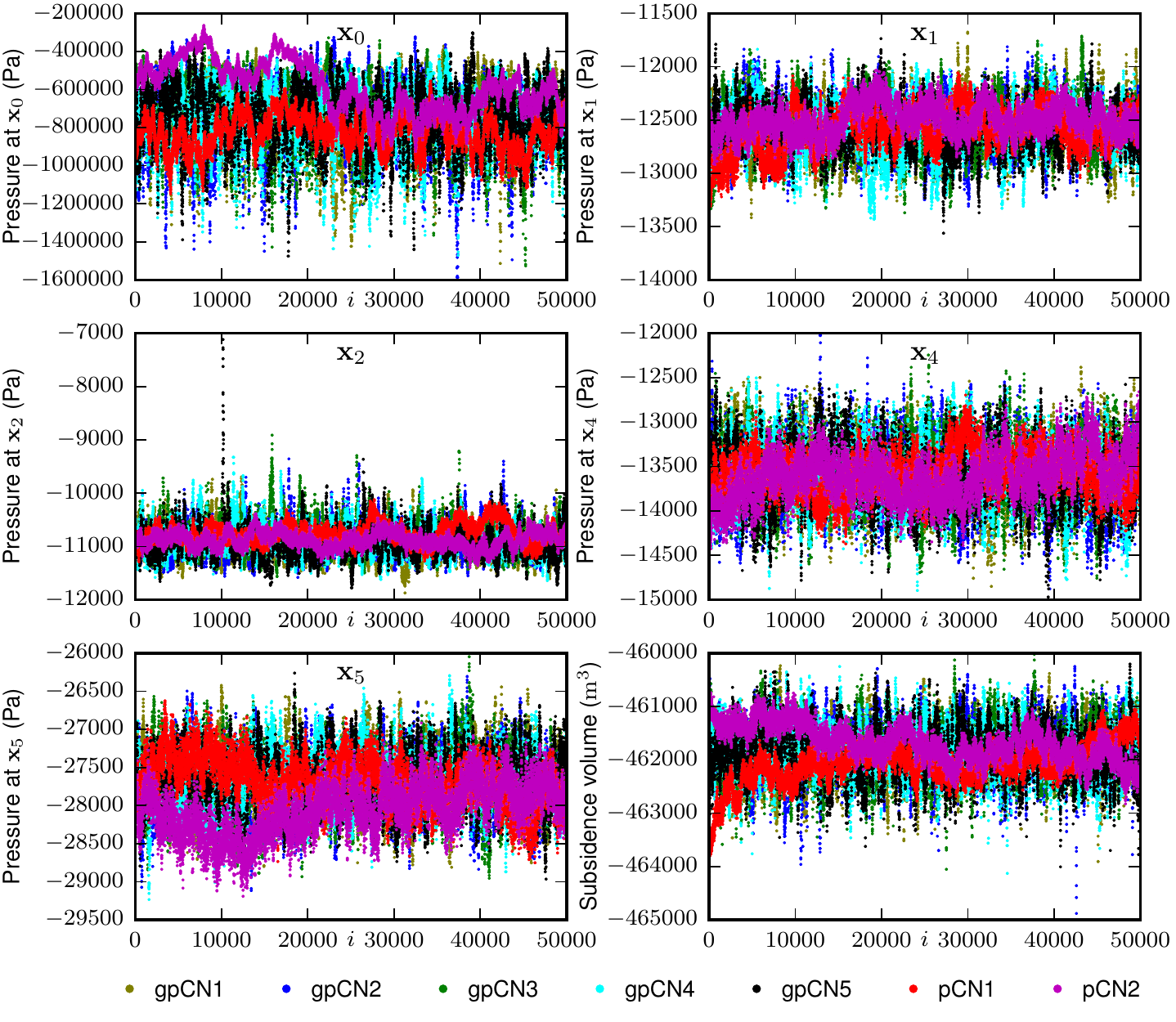}
\caption
{pCN and gpCN MCMC-based mixing of the pointwise pressure values at selected locations ($\mathbf{x}_0, \mathbf{x}_1, \mathbf{x}_2, \mathbf{x}_4$, and $\mathbf{x}_5$) and total subsidence volume (bottom right panel). The two pCN chains and the five gpCN chains are labeled pCN1, pCN2, and  gpCN1--gpCN5, respectively.}
\label{fig: Mixing for State}
\end{figure}

Quantifying the propagation of  uncertainty in the parameters to the state variables QoIs, e.g.\ pore pressure  and total subsidence, is of interest in GW management applications \cite{QinAndrewsFang2018}. We show the prior and posterior predictive distributions of the pressure $p(\mathbf{x}_i,t_\text{InSAR})$, at the well location $\mathbf{x}_0=(0,0,.6)$, and locations $\mathbf{x}_1, \mathbf{x}_2, \mathbf{x}_4$ and  $\mathbf{x}_4$ (coordinates are provided in Figure~\ref{fig: Parameter Mixing}) in Figure~\ref{fig: QoI Marginals}. We also show these predictive distributions for total subsidence volume $-\int_{\Gamma^\text{top}} u_3(t_\text{InSAR}) ds$, where $\Gamma^\text{top}$ is the domain top surface (bottom panel in Figure~\ref{fig: QoI Marginals}). We use KDE to estimate the PDF of each of those QoIs.
We note that these distributions can be highly non-Gaussian and that the prior-based distributions vary across a much wider range compared to the narrow posterior-based distributions. In some locations, $\mathbf{x}_1$, $\mathbf{x}_2$ and $\mathbf{x}_5$ for example, relying only on prior knowledge significantly underestimates the expected pressure drop; the posterior-based QoI mean at these locations is considerably larger in magnitude than that of the prior-based QoI. We also note appreciable differences between the low-rank based Laplace approximation and the MCMC based posterior predictives. The mixing behaviour of the chains for these QoIs distributions (shown in Figure~\ref{fig: Mixing for State}) is similar to what was observed for the parameter chains in section~\ref{sec: gpCN Performance}.

%%%%%%%%%%%%%%%%%%%%%%%%%%%%%%%%%%%%%%%%%%%%%%%%%%%%%%%%%%%%%%%%%%%%%%%%%%%%%%%
%%%%%%%%%%%%%%%%%%%%%%%%%%% Section: Conclusion  %%%%%%%%%%%%%%%%%%%%%%%%%%%%%%%
%%%%%%%%%%%%%%%%%%%%%%%%%%%%%%%%%%%%%%%%%%%%%%%%%%%%%%%%%%%%%%%%%%%%%%%%%%%%%%%
\section{Conclusion}
\label{sec: Conclusion}

The use of InSAR surface deformation data to quantify the uncertainty in GW aquifer models via solution of poroelasticity-governed high dimensional Bayesian inverse problems is important for data-driven, model-predictive management of GW resources. The intrinsic low dimensionality of this inverse problem, evident by the rapid decay in data misfit Hessian eigenfunctions, makes solving this problem computationally tractable when the right computational tools are used, e.g.\ our proposed gpCN MCMC method employing low-rank-based Laplace approximation proposals, scalable priors based on inverses of differential operators, and an inexact Newton method with adjoint-based gradients and Hessian actions to find the MAP point. The use of these tools guarantees that the number of forward poroelasticity PDE solves required to solve the inverse problem is independent of the parameter dimension. In the Nevada test case, for example, only $r\approx 80$ (out of $n_n=4{,}081$) directions are sufficient to construct the low-rank based Laplace approximation.

 The information content of InSAR data is rich enough to enable characterizing the permeability and state variable derived QoI with high certainty. Location-wise, the permeability in the aquifer in regions that undergo a detectable subsidence signal or large subsidence gradient is inferred with narrow variance. However, what can be inferred about areas that are ``shadowed'' from the source  (well WX31) beyond a flow barrier is very limited. The update in the prior statistical characterization of selected state variable derived QoI via learning from InSAR data is also critical. Our results show for example that relying on prior knowledge alone leads to underestimating the pressure drop in some locations.

%%%%%%%%%%%%%%%%%%%%%%%%%%%%%%%%%%%%%%%%%%%%%%%%%%%%%%%%%%%%%%%%%%%%%%%%%%%%%%%
%%%%%%%%%%%%%%%%%%%%%%%%%%% Section: Appendices   %%%%%%%%%%%%%%%%%%%%%%%%%%%%%%%
%%%%%%%%%%%%%%%%%%%%%%%%%%%%%%%%%%%%%%%%%%%%%%%%%%%%%%%%%%%%%%%%%%%%%%%%%%%%%%%
\FloatBarrier
\newpage
\appendix

%%%%%%%%%%%%%%%%%%%%%%%%%%%%%%%%%%%%%%%%%%%%%%%%%%%%%%%%%%%%%%%%%%%%%%%%%%%%%%
%%%%%%%%%%%%%%%%%%%%%%%%%%%%%%%%%%%%%%%%%%%%%%%%%%%%%%%%%%%%%%%%%%%%%%%%%%%%%%
\section{Adjoint-Based Derivation of the Hessian Operator}
%\index{Background II@\emph{Background II}}%
\label{app: Hessian}

%========================================================================

We derive the Hessian operator in the Newton iteration that we form for finding the MAP point. Maximizing the posterior distribution \eqref{equ: Bayes Formula Finite} with respect to the parameter $\mathbf{m}$ to find the MAP point is equivalent to minimizing the negative log posterior, i.e.\
\begin{align}
\min_{\mathbf{m}\in {\rm I\!R}^{n_n}} \mathcal{J}(\mathbf{m}) = 
\frac{1}{2}(\gt{B}\gt{X} - &\mathbf{d}^{\text{obs}}  )^T
\mathbf{\Gamma}_\text{noise}^{-1}( \gt{B}\gt{X} -\mathbf{d}^{\text{obs}})
 +  \frac{1}{2}(\mathbf{m} - \bar{\mathbf{m}})^T \mathbf{\Gamma}_\text {prior}^{-1}
 (\mathbf{m} - \bar{\mathbf{m}}),
 \label{equ: Constrained Minimization}
\end{align}
where $\gt{X}$ depends on $\mathbf{m}$ through the solution of the forward problem $\gt{S}(\mathbf{m})\gt{X} = \gt{F}$.

To derive the gradient using the adjoint method, we form the Lagrangian for the constrained optimization problem~\eqref{equ: Constrained Minimization} to be:
\begin{align}
 \mathcal{L}(\mathbf{m},\gt{X},\gt{Y}) = 
&\frac{1}{2}(\gt{B}\gt{X} - \mathbf{d}^{\text{obs}}  )^T
\mathbf{\Gamma}_\text{noise}^{-1}( \gt{B}\gt{X}
-\mathbf{d}^{\text{obs}})\nonumber  \\
+&  \frac{1}{2}(\mathbf{m} - \bar{\mathbf{m}})^T\mathbf{\Gamma}_\text{prior}^{-1}
 (\mathbf{m} - \bar{\mathbf{m}}) 
 + \gt{Y}^T(\gt{S}(\mathbf{m})\gt{X}- \gt{F}).   \label{equ: First Lagrangian}
\end{align}
where  $\gt{Y}$ is the Lagrange multiplier (also called the adjoint variable).

At a minimum of \eqref{equ: First Lagrangian}, the partial derivatives of the Lagrangian with respect to the state variable, ${\partial\mathcal{L}}/{\partial \gt{X}}$, the parameter, ${\partial\mathcal{L}}/{\partial \mathbf{m}}$, and  the adjoint variable, ${\partial\mathcal{L}}/{\partial \gt{Y}}$, vanish. We set  ${\partial\mathcal{L}}/{\partial \gt{X}}$ and  ${\partial\mathcal{L}}/{\partial \gt{Y}}$ to zero, which gives the state equation and the adjoint equation respectively:
\begin{align}
\frac{\partial\mathcal{L}}{\partial \gt{Y}}(\gt{X}, \gt{Y}, \mathbf{m}) =& \gt{S}(\mathbf{m})\gt{X} -
\gt{F} =
0  \quad \text{(state)},
\label{equ: State Equation}\\
\frac{\partial\mathcal{L}}{\partial \gt{X}}(\gt{X}, \gt{Y}, \mathbf{m})   =&
\gt{B}^T\mathbf{\Gamma}_\text{noise}^{-1} ( \gt{B}\gt{X} -
\mathbf{d}^{\text{obs}}  ) + \gt{S}^T(\mathbf{m})\gt{Y} = 0 \quad \text{(adjoint)}.
\label{equ: Adjoint Equation}
\end{align}
We satisfy these two conditions by solving the discretized state and adjoint equations exactly for the given value of $\mathbf{m}$. Thus we seek the parameter $\mathbf{m}$ that ensures that the gradient
 \begin{align}
   \frac{\partial\mathcal{L}}{\partial \mathbf{m}}(\gt{X}, \gt{Y}, \mathbf{m}) =& \mathbf{\Gamma}_\text{prior}^{-1}(\mathbf{m} -
\bar{\mathbf{m}}) + \gt{C}^T(\mathbf{m})\gt{Y}
\label{equ: Gradiant}
\end{align}
vanishes. The operator $\gt{C}$ is the partial derivative (sensitivity) of the residual $\mathcal{R} = \gt{S}\gt{X} - \gt{F}$ with respect to the parameter $\mathbf{m}$
\begin{align}
\gt{C} = \frac{\partial }{\partial \mathbf{m}} (\gt{S}\gt{X} - \gt{F}) =
\frac{\partial }{\partial \mathbf{m}} (\gt{S}\gt{X} ).
\label{equ: sensitivity}
\end{align}
We solve the equation ${\partial\mathcal{L}}/{\partial \mathbf{m}} =0$ using Newton's method.

To derive the Newton iteration we form the Lagrangian $\mathcal{L}_\mathcal{H}$:
\begin{align}
 \mathcal{L}_\mathcal{H}:= & {\delta\gt{X}}^T \left(
\gt{B}^T\mathbf{\Gamma}_\text{noise}^{-1} ( \gt{B}\gt{X} - \mathbf{d}^{\text{obs}}  ) 
  + \gt{S}^T(\mathbf{m})\gt{Y}\right)\nonumber\\
 + &{\delta\gt{Y}}^T \big( \gt{S}(\mathbf{m})\gt{X} - \gt{F}\big)\nonumber\\
 + &{\delta \mathbf{m}}^T \big( \mathbf{\Gamma}_\text{prior}^{-1} (\mathbf{m} - \bar{\mathbf{m}}) + \gt{C}^T(\mathbf{m})\gt{Y}\big),
\end{align}
where $\delta\gt{X}$ and $\delta\gt{Y}$ are Lagrange multipliers for the adjoint \eqref{equ: Adjoint Equation} and state \eqref{equ: State Equation} equations, and $\delta\mathbf{m}$ is the direction in which the Hessian acts. The so-called incremental forward and incremental adjoint problems are defined as follows:
\begin{align}
\frac{\partial\mathcal{L}_\mathcal{H}}{\partial \gt{Y}}  &= \gt{S}(\mathbf{m}){\delta\gt{X}} +
\gt{C}(\mathbf{m}){\delta\mathbf{m}} =0. \quad \text{(incremental forward)},\label{equ: Incremental Forward}\\
\frac{\partial\mathcal{L}_\mathcal{H}}{\partial \gt{X}} &=
\gt{B}^T\mathbf{\Gamma}_\text{noise}^{-1}\gt{B}  {\delta\gt{X}} + 
\gt{S}^T(\mathbf{m}){\delta\gt{Y}} +\Big( \frac{\partial}{\partial \gt{X}} (\gt{C}^T(\mathbf{m})\gt{Y})\Big)^T\delta \mathbf{m}  \nonumber \\
&=  \mathbf{W}_{\gt{X}\gt{X}}  {\delta\gt{X}} 
 + \gt{S}^T(\mathbf{m}){\delta\gt{Y}} + \mathbf{W}_{\gt{X}\mathbf{m}}(\mathbf{m})\delta \mathbf{m} =
0.\quad \text{(incremental adjoint)}\label{equ: Incremental Adjoint}
\end{align}

The Hessian action in the direction $\delta\mathbf{m}$ is given by:
\begin{align}
  \frac{\partial\mathcal{L}_\mathcal{H}}{\partial \mathbf{m}}
&=   \frac{\partial}{\partial \mathbf{m}} ( {\delta\gt{X}}^T\gt{S}^T(\mathbf{m})\gt{Y})
+  \frac{\partial}{\partial \mathbf{m}} ( {\delta\gt{Y}}^T\gt{S}(\mathbf{m})\gt{X})
+    \mathbf{\Gamma}_\text{prior}^{-1}\delta\mathbf{m} + \frac{\partial}{\partial \mathbf{m}}  ({\delta
\mathbf{m}}^T \gt{C}^T(\mathbf{m})\gt{Y})\nonumber\\
&=  \mathbf{W}_{\mathbf{m}\gt{X}}(\mathbf{m}) \delta\gt{X} +  \gt{C}^T(\mathbf{m}) \delta\gt{Y}
+  \mathbf{\Gamma}_\text{prior}^{-1}\delta\mathbf{m} +
\mathbf{R}_{\mathbf{m}\mathbf{m}}(\mathbf{m})\delta\mathbf{m}. \label{equ: Hessian Expression} 
\end{align}
Substituting the solutions for $\delta\gt{X}$ and $\delta\gt{Y}$ from the systems~\eqref{equ: Incremental Forward} and \eqref{equ: Incremental Adjoint} into the expression~\eqref{equ: Hessian Expression} gives the Hessian expression:
\begin{align}
 \mathbf{H}(\mathbf{m}) = &
 \mathbf{\Gamma}_\text{prior}^{-1} + \mathbf{H}_\text{misfit}(\mathbf{m}),
 \label{equ: Full Hessian}
 \end{align}
 where
\begin{align}
\mathbf{H}_\text{misfit}(\mathbf{m}) \coloneqq & \mathbf{R}_{\mathbf{m} \mathbf{m}}(\mathbf{m}) +\gt{C}^{T}(\mathbf{m}) \gt{S}^{-T}(\mathbf{m}) (
\mathbf{W}_{\gt{X}\gt{X}} \gt{S}^{-1}(\mathbf{m}) \gt{C}(\mathbf{m})-   \mathbf{W}_{\gt{X}\mathbf{m}}(\mathbf{m}) ) \nonumber\\
 &-
\mathbf{W}_{\mathbf{m} \gt{X}}(\mathbf{m})\gt{S}^{-1}(\mathbf{m})\gt{C}(\mathbf{m}).
\label{equ: Data Misfit Hessian}
\end{align}
The matrices $\mathbf{W}_{\gt{X}\gt{X}}, \mathbf{W}_{\gt{X}\mathbf{m}}, \mathbf{W}_{\mathbf{m} \gt{X}}$, and $\mathbf{R}_{\mathbf{m} \mathbf{m}}$ are defined as follows 
\begin{align}
\mathbf{W}_{\gt{X}\gt{X}} = \gt{B}^T\mathbf{\Gamma}_\text{noise}^{-1}\gt{B},\ \mathbf{W}_{\gt{X}\mathbf{m}}= \frac{\partial}{\partial\mathbf{m}}(\gt{S}^{T}\gt{Y}),\ \mathbf{W}_{\mathbf{m} \gt{X}} = \frac{\partial}{\partial \gt{X}}(\gt{C}^{T}\gt{Y}),\ \mathbf{R}_{\mathbf{m} \mathbf{m}}=   \frac{\partial}{\partial \mathbf{m}} (\gt{C}^{T}\gt{Y}).\nonumber
\end{align}

Forming the Hessian, evaluated at a given $\mathbf{m}$, explicitly, requires applying the operator $\gt{S}^{-1}$ to each column of  $\gt{C}$ to form the matrices product $\gt{S}^{-1} \gt{C}$. Applying the operator $\gt{S}^{-1}$ to a vector amounts to an incremental forward poroelasticity solve, equation~\eqref{equ: Incremental Forward}. Therefore, forming the GN Hessian explicitly costs $n_n$ poroelasticity PDEs solves (note that no additional solves are required for forming the product $\gt{C}^{T} \gt{S}^{-T}$ since it is the transpose of the product $\gt{S}^{-1} \gt{C}$). However, applying  $\mathbf{H}$ to a direction mainly costs applying the operators $\gt{S}^{-1}$ and $\gt{S}^{-T}$ each to a vector which is the cost of two poroelasticity PDEs solves (one incremental forward~\eqref{equ: Incremental Forward} and one incremental adjoint~\eqref{equ: Incremental Adjoint}).

\section{Forward Model Parameters}
\label{app: Forward Model Parameters}

%%%%%%%%%%%%%%%%%%%%%%% Parameters of the model
\begin{table}[h]
\caption[The forward Model Parameters]{Forward model parameters} 
\begin{center}
\subcaption{Model parameters}
\begin{tabular}{ c|c }
\hline
 Volumetric force ($\mathbf{f_u}$)  &  0    \\
 Pumping rate ($-\int_\Omega f_p \mathrm{d}\mathbf{x}$)          &  $9{,}028$ $ \text{m}^3/ \text{day}$   \\
 Fluid viscosity ($\mu$)          &  $0.001$ $ \text{Pa}\cdot\text{s} $   \\
Biot-Willis coefficient ($\alpha$)       &  $0.998 $   \\
Height of the domain ($h$)            &  $885$ $\text{m} $  \\
Water density ($\rho$)         &  $997.97$ $\text{Kg}/\text{m}^3$ \\
Gravitational acceleration ($g$)            &  $ 9.8$ $\text{m}/\text{s}^2$ \\
Water compressibility  ($\beta_w$)      &  $ 4.4 \times10^{-10}$ $\text{Pa}^{-1}$ \\
 \hline
\end{tabular}\\[0.3in]
\subcaption{Layer parameters}
\begin{tabular}{ c|c|c|c|c }
\hline
                                             &  Units   & Aquifer                    & Confining layer        &  Lower clay          \\
\hline
Poisson's ratio ($\nu$)                      &    -            & $0.25$                     &  $0.25$                & $0.25$               \\
Drained shear modulus ($G$)                  &    Pa           & $3.4 \times 10^8$          & $3.5 \times 10^8$      & $8 \times 10^8$      \\ 
Specific storage ($S_\epsilon$)              & $\text{Pa}^{-1}$   & $2.1 \times 10^{-9}$       &  $1.2 \times 10^{-9}$  & $1.2 \times 10^{-9}$ \\ 
Prior mean value ($\text{log}_{10}(\kappa)$) &  $\text{m}^{2}$ (for $\kappa$)      &     -11.2                  &         -14.2          &     -14.2                  \\
%Hight of each unit($h$)               & 400 m            & 85 m          & 400 m              \\
\hline 
\end{tabular}
\end{center}
\label{tab: Biot Model Parameters}
\end{table}

\FloatBarrier
\acknowledgments

This work was supported by National Science Foundation (NSF) Grants CBET–1508713 and ACI-1550593, and DOE grant DE-SC0019303. A.A. would like to acknowledge funding from the Ministry of Education in Saudi Arabia. The Envisat SAR imagery
used in this study can be downloaded through the UNAVCO Data Center SAR archive.

%% ------------------------------------------------------------------------ %%
%% References and Citations

%%%%%%%%%%%%%%%%%%%%%%%%%%%%%%%%%%%%%%%%%%%%%%%
%
% \bibliography{<name of your .bib file>} don't specify the file extension
%
% don't specify bibliographystyle
%%%%%%%%%%%%%%%%%%%%%%%%%%%%%%%%%%%%%%%%%%%%%%%
%\bibliography{amal,references,marc,insar}

\end{document}